\documentclass[pageno]{jpaper}    

\usepackage{afterpage}
\usepackage{graphicx}
\usepackage{makecell}
\usepackage{xcolor}
\usepackage{dingbat}
\usepackage{multirow}
\usepackage{amssymb}
\usepackage{listings}
\usepackage{pifont}
\usepackage{threeparttable}
\usepackage{caption}
\usepackage{fancyhdr}
\usepackage{comment}

\definecolor{mGreen}{rgb}{0,0.6,0}
\definecolor{mGray}{rgb}{0.5,0.5,0.5}
\definecolor{mPurple}{rgb}{0.58,0,0.82}
\definecolor{backgroundColour}{rgb}{0.95,0.95,0.95}

\lstdefinestyle{CStyle}{
    backgroundcolor=\color{backgroundColour},   
    commentstyle=\color{mGreen},
    keywordstyle=\color{blue},
    stringstyle=\color{mPurple},
    basicstyle=\footnotesize,
    breakatwhitespace=false,         
    breaklines=true,                 
    captionpos=b,                    
    keepspaces=true,                 
    numbersep=5pt,                  
    showspaces=false,                
    showstringspaces=false,
    showtabs=false,                  
    tabsize=2,
    language=C
}

\newcommand{\squishlist}{
 \begin{list}{$\bullet$}
  { \setlength{\itemsep}{0pt}
     \setlength{\parsep}{3pt}
     \setlength{\topsep}{3pt}
     \setlength{\partopsep}{0pt}
     \setlength{\leftmargin}{1.5em}
     \setlength{\labelwidth}{1em}
     \setlength{\labelsep}{0.5em} } }

\newcommand{\squishlisttwo}{
 \begin{list}{$\bullet$}
  { \setlength{\itemsep}{0pt}
     \setlength{\parsep}{0pt}
    \setlength{\topsep}{0pt}
    \setlength{\partopsep}{0pt}
    \setlength{\leftmargin}{2em}
    \setlength{\labelwidth}{1.5em}
    \setlength{\labelsep}{0.5em} } }

\newcommand{\squishend}{
  \end{list}  }

\newcommand{\ignore}[1]{}

\usepackage{xcolor, soul}
\newcommand{\greencheck}{{\color{green}\checkmark}}

\newcommand{\redcheck}{{\color{red}\xmark}}

\newcommand{\xmark}{\ding{55}}%
\newcommand{\jl}[1]{{\color{blue}\bfseries [JL::: #1]}}
\newcommand{\ma}[1]{{\color{green}\bfseries [Manuel::: #1]}}
\newcommand{\pa}[1]{{\color{orange}\bfseries [Fran::: #1]}}
\newcommand{\TK}[1]{{\color{teal}\bfseries [TK::: #1]}}
\newcommand{\RG}[1]{{\color{cyan}\bfseries [RG::: #1]}}
\newcommand{\MP}[1]{{\color{purple}\bfseries [Michael::: #1]}}

\definecolor{aquamarine}{rgb}{0.5, 1.0, 0.83}
\definecolor{ashgrey}{rgb}{0.7, 0.75, 0.71}
\definecolor{atomictangerine}{rgb}{1.0, 0.6, 0.4}
\definecolor{babyblue}{rgb}{0.54, 0.81, 0.94}
\definecolor{fluorescentyellow}{rgb}{0.8, 1.0, 0.0}
\definecolor{lavender(floral)}{rgb}{0.71, 0.49, 0.86}
\definecolor{mauve}{rgb}{0.88, 0.69, 1.0}

\newcommand{\AcceleratorName}{{Flexagon}}
\newcommand{\SramCName}{{PSRAM }}

\newcommand{\innerprod}{\texttt{IP}}
\newcommand{\outerprod}{\texttt{OP}}
\newcommand{\gustavsons}{\texttt{Gust}}

\newcommand{\innerprodM}{\texttt{IP(M)}}
\newcommand{\outerprodM}{\texttt{OP(M)}}
\newcommand{\gustavsonsM}{\texttt{Gust(M)}}

\newcommand{\innerprodN}{\texttt{IP(N)}}
\newcommand{\outerprodN}{\texttt{OP(N)}}
\newcommand{\gustavsonsN}{\texttt{Gust(N)}}

\begin{document}
\date{}
\title{\AcceleratorName: A Multi-Dataflow Sparse-Sparse Matrix Multiplication Accelerator for Efficient DNN Processing}

\newcommand{\changed}[1]{{{\color{red} #1}}}

\author{
Francisco Mu\~noz-Mart\'{i}nez$^*$~~~~~~~~~~~~~~~~~~~~~~~~~~~~~~Raveesh Garg$^\dagger$~~~~~~~~~~~~~~~~~~~~~~~~~~~Jos\'{e}~L.~Abell\'{a}n$^{*}$\\
Manuel~E.~Acacio$^*$~~~~~~~~~~~~~~~~~~~~~~~~~~~~~~~~~~~~~~~~~~~Tushar Krishna$^\dagger$~~~~~~~~~~~~~~~~~~~~~~~~~~Michael Pellauer$^\ddagger$ \\
\\
$^*$Universidad de Murcia,~~~~$^\dagger$Georgia Tech,~~~~$^\ddagger$NVIDIA \\
\\
Contact author: \texttt{francisco.munoz2@um.es}
}






\maketitle

\pagestyle{plain}


\begin{abstract}

Sparsity is a growing trend in modern DNN models.
Existing Sparse-Sparse Matrix Multiplication (SpMSpM) accelerators are tailored to a particular SpMSpM dataflow (i.e., Inner Product, Outer Product or Gustavson's), that determines their overall efficiency. We demonstrate that this static decision inherently results in a suboptimal dynamic solution. This is because different SpMSpM kernels show varying features (i.e., dimensions, sparsity pattern, sparsity degree), which makes each dataflow better suited to different data sets. 

In this work we present Flexagon, the first SpMSpM reconfigurable accelerator that is capable of performing SpMSpM computation by using the particular dataflow 
that best matches each case. Flexagon accelerator is based on a novel Merger-Reduction Network (MRN) that unifies the concept of reducing and merging in the same substrate, increasing efficiency. Additionally, Flexagon also includes a 3-tier memory hierarchy, specifically tailored to the different access characteristics of the input and output compressed matrices. Using detailed cycle-level simulation of contemporary DNN models from a variety of application domains, we show that Flexagon achieves average performance benefits of 4.59$\times$, 1.71$\times$, and 1.35$\times$ with respect to the state-of-the-art SIGMA-like, Sparch-like and GAMMA-like accelerators (265\% , 67\% and 18\%, respectively, in terms of average performance/area efficiency).

\end{abstract}

\section{Introduction}   
\label{section:introduction}


Sparsity in tensors is an emerging trend 
in modern DNN workloads~\cite{reddi2020mlperf, mattson2019mlperf,DLRM19}. These workloads have diverse sparsity ratios, ranging from 
0.04\% to 90\%, and are used in various applications, ranging from personalized recommendations~\cite{DLRM19} to Natural Language Processing~\cite{BERT2020}. Sparsity in weights stems from pruning ~\cite{Han2016} and sparsity inside activations stems from nonlinear functions such as ReLU. As a result, exploiting the benefits of sparsity by directly implementing sparse matrix-matrix multiplication (SpMSpM) has become an important target for customized DNN accelerators~\cite{SIGMA2020,hegde2019extensor,SpArch2020,outerspace,zhang2021gamma,srivastava2020matraptor,SCNN2017}. 

The most common way for these accelerators to exploit sparsity is using compressed formats like Bitmap, CSR and CSC to store and operate (multiply and accumulate) only the non-zero values. This allows to significantly reduce both the memory footprint and the number of operations, which in turn translates into significant energy savings. However, these accelerators vary widely in their hardware implementation and in the exploited dataflow. The dataflows used by these accelerators in terms of the loop order of computation have been broadly classified into \texttt{Inner Product} (\innerprod), \texttt{Outer Product} (\outerprod)~and Row-wise-Product, often called \texttt{Gustavson's} (\gustavsons)~\cite{gustavson:fastMM}. 

\begin{table}[t!]
\begin{footnotesize}
\begin{center}
{
\begin{tabular}{|c|c|c|c|c|}\hline
 \textbf{Accelerator} & \textbf{Architectural Features} & \textbf{IP} & \textbf{OP} & \textbf{Gust} \\\hline
 TPU~\cite{jouppi-2017} & Dense Systolic Array  & N/A & N/A  & N/A \\\hline
 SIGMA~\cite{SIGMA2020} & Configurable Reduce Tree & \greencheck & \redcheck  & \redcheck
 \\\hline
 ExTensor~\cite{hegde2019extensor} & Intersection Unit & \greencheck & \redcheck  & \redcheck
 \\\hline
 MatRaptor~\cite{srivastava2020matraptor} & Merger & \redcheck & \redcheck  & \greencheck
 \\\hline
 Gamma~\cite{zhang2021gamma} & Fiber Cache, Merger & \redcheck & \redcheck  & \greencheck \\\hline
 Outerspace~\cite{outerspace} & Merger & \redcheck & \greencheck  & \redcheck\\\hline
 SpArch~\cite{SpArch2020} & Matrix condenser, merger  & \redcheck & \greencheck  & \redcheck
 \\\hline
 Flexagon & Flexible Merge/Reduce & \greencheck & \greencheck  & \greencheck
 \\
 (ours) & tree and memory controller & & &
 \\\hline

\end{tabular}
}
\end{center}
\end{footnotesize}
\vspace{-0.45cm}
\caption{Comparison of Flexagon with prior Sparse DNN accelerators in terms of supported dataflows. 
IP=Inner Product, OP=Outer Product, Gust=Gustavson's (Row-wise Product).}
\label{table:related}
\vspace{-0.15cm}
\end{table}

 \begin{figure}[t!]
        \begin{center}
                \includegraphics[width=0.8\columnwidth]{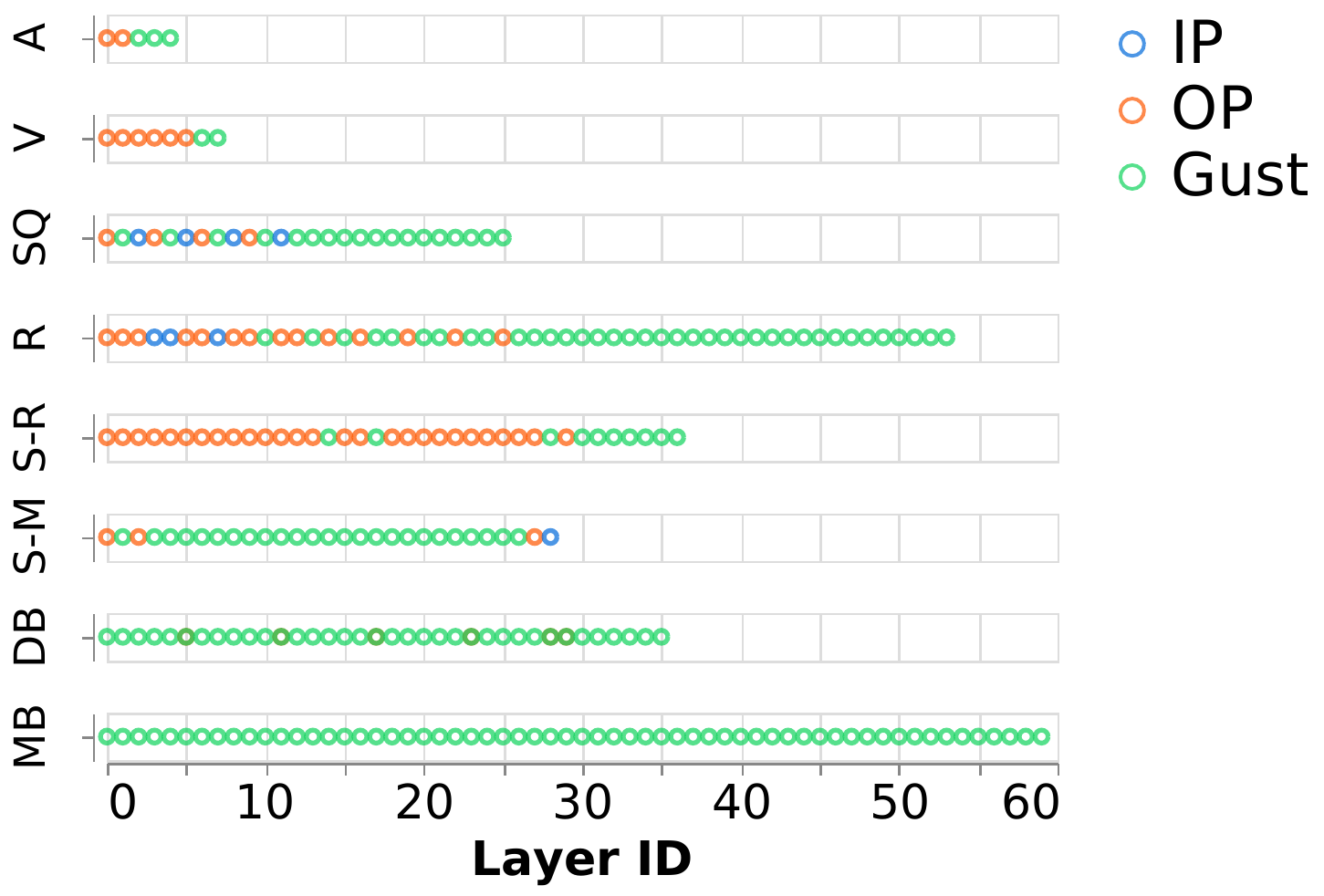}
        \end{center}
        \vspace{-0.40cm}
        \caption{Dataflow that obtains the best performance per layer across the DNN models (see Table~\ref{table:dnn_models} that includes their sparsity ratios). 
        IP=\textit{Inner Product}, OP=\textit{Outer Product} and Gust=\textit{Gustavson's}. 
        }
        \label{fig:layers_dataflow}
        \vspace{-0.1cm}
\end{figure}

Table~\ref{table:related} shows prior sparse accelerators and the dataflows they support.
While state-of-the-art sparse accelerators such as SIGMA~\cite{SIGMA2020}, Sparch~\cite{SpArch2020} and GAMMA~\cite{zhang2021gamma} have been optimized for a fixed dataflow (\innerprod, \outerprod~and~     
\gustavsons, respectively), in this paper, we make the important observation that \textit{the optimal dataflow changes from a DNN model to another, and even within a DNN model, from one layer to another}, so that contemporary fixed-dataflow accelerators cannot adapt well to maximize DNN application performance.

To back up our observation, Fig.~\ref{fig:layers_dataflow} shows the dataflow that obtains the best performance per layer given the execution  of 8 entire DNN models obtained from MLPerf benchmark suite~\cite{reddi2020mlperf} as well as some extra models (details in Table~\ref{table:dnn_models}).  Observe that we consider heterogeneous models from different domains, sizes  and sparsity ratios. For \textit{MB}, we only show the first 60 layers, which represent 20\% out of the total number of layers. 
To model the three dataflows, the executions have been performed on a 64-Multiplier SIGMA-like, Sparch-like and GAMMA-like architectures (further details in Section~\ref{section:methodology}). 
The NLP models \textit{DB} and \textit{MB} present a clear trend towards \gustavsons. 
On the other hand, extremely sparse models, such as \textit{S-R} and \textit{V}, benefit from \outerprod~in 73\% and 75\% of the layers, respectively. The rest of the DNN models  present a high variability across layers, and the most efficient dataflow changes given the different features of each layer. This highlights that one dataflow does not fit all, and so there is an opportunity to increase efficiency via dynamic adaptation of the architectural components to the most suitable dataflow.

\begin{table*}[t!]
\begin{footnotesize}
\begin{center}
{
\begin{tabular}{|c|c|c|c|c|c|c|c|c|c|c|c|}\hline
 \textbf{DNN} 
 & \textbf{Appl} & \textbf{nl} & \textbf{AvSpA} & \textbf{AvSpB} & \textbf{AvCsA} & \textbf{AvCsB} & {\textbf{MinCsA}} & {\textbf{MinCsB}} & \textbf{MaxCsA} & \{\textbf{MaxCsB} & {\textbf{Cycles($10^6$) CPU}}  \\\hline
 \textit{Alexnet (A)} & CV & 7 &  70  & 48 & 0.56 & 13.6 & {0.02} & {0.18} & {1.01} & {63.41} & {3804} \\\hline
 \textit{Squeezenet (S)} & CV & 26 & 70   & 31 & 0.05 & 1.54 & {0.001} & {0.02} & {0.58} & {26.6} & {2751} \\\hline
 \textit{VGG-16 (V)} & CV & 8 & 90 & 80  & 0.55 & 2.90 & {0.02} & {0.15} & {10.42} & {0.90} & {6012} \\\hline
 \textit{Resnets-50 (R)}  & CV & 54 & 89 & 52 & 0.19  & 1.30 & {0.001} & {0.007} & {1.0} & {26.64} & {4185} \\\hline
 \textit{SSD-Resnets (S-R)} & OR & 37 & 89  & 49  & 0.12 & 3.60 & {0.003} & {0.003} & {10.1} & {0.50} & {6429} \\\hline
 \textit{SSD-Mobilenets (S-M)}  &  OR & 29 & 74 & 35 & 0.16 & 0.31 & {0.002} & {0.0004} & {1.0} & {1.65} &  {5379} \\\hline
 \textit{DistilBERT (DB)} & NLP  & 36 & 50  & 0.04  & 2.25  & 0.35 & {1.12} & {0.23} & {4.5} & {0.94} &  {5748} \\\hline
 \textit{MobileBERT (MB)} & NLP & 316  & 50 & 11 & 0.10 & 0.07 & {0.03} & {0.003} & {0.125} & {0.01} & {4893} \\\hline
\end{tabular}
}
\end{center}
\end{footnotesize}
\vspace{-0.25cm}
\caption{DNN models used in this work. Appl=Application domain. CV=Computer Vision, OR=Object Recognition, NLP=Natural Language Processing, nl=Number of layers, AvSp\{A,B\}=Average sparsity of the matrices \{A,B\} (in \%), AvCs\{A,B\}=Average compressed matrix size for the matrices \{A,B\} (in MiB). MinCs\{A,B\}=Minimum compressed matrix size for the matrices \{A,B\} (in MiB). MaxCs\{A,B\}=Maximum compressed matrix size for the matrices \{A,B\} (in MiB). Cycles($10^6$) CPU = Number of cycles obtained after running the benchmarks using MKL in a CPU system.} 
\label{table:dnn_models}
\vspace{-0.25cm}
\end{table*}

The value of supporting flexible dataflows has been explored extensively for dense DNNs~\cite{MAESTRO2019, TimeLoop2019,MAERI2018,eyerissv2}.
However, support for flexible dataflow acceleration for sparse workloads is much more challenging 
because of different ways in which these accelerators handle sparsity. For example, the \innerprod~dataflow implemented in SIGMA~\cite{SIGMA2020} implements a reduction network called FAN to \textit{reduce} the generated partial sums at once, as well as the capacity to perform \textit{intersections} to execute a sparse dot product. On the contrary, the \outerprod~and~\gustavsons~ dataflows implemented in accelerators like Sparch~\cite{SpArch2020} and GAMMA~\cite{zhang2021gamma} produce partial sums instead of complete sums, and hence, require \textit{merging} the non-zero partial sums and use merger trees for this purpose.
A naive implementation using separate hardware widgets for reductions and merges 
would lead to significant area overhead (see Section~\ref{section:rtl_results}).

To efficiently support different SpMSpM workloads to run modern sparse DNNs, we present \textit{Flexagon}, the first (to our knowledge) reconfigurable sparse and homogeneous DNN accelerator that can be dynamically adapted to execute the most suited SpMSpM dataflow on a per DNN layer basis. 
Flexagon features a novel unified \textit{Merger-Reduction Network (MRN)} that supports both reduction of dot products and merging of partial sums.  We propose a tree-based topology where the nodes are configured to act either as accumulators or comparators, as explained in~Section~\ref{section:design}. 
Flexagon also features a new L1 on-chip memory organization composed of three customized memory structures that are able to capture the memory access pattern of each dataflow. 
The first memory structure is a simple read-only FIFO, which is designed for the sequential accesses that occur during some stages in the three dataflows. The second one is a low-power cache used to back-up the random accesses caused mainly by the \gustavsons~dataflow.  Finally, a customized memory structure called \SramCName is specifically designed to store and read psums, which is essential for both~\outerprod~ and~\gustavsons~ dataflows. These memory structures allow us to support all the three dataflows with minimal area and power overheads.  
Further, 
our accelerator also prevents the hardware from requiring explicit expensive conversions of compression formats (i.e., from CSR to CSC or vice-versa)~\cite{qin2021extending} between layers as it is possible to easily switch among the most convenient dataflow given a particular compression format (details discussed in Section~\ref{section:design}).



We summarize our key contributions:

(1) We demonstrate that each SpMSpM operation in modern sparse DNN layers presents different memory 
access patterns according to matrix dimensions and sparsity patterns. As a consequence, the dataflow 
that maximizes the performance of a particular SpMSpM operation not only can change between DNN models, but also from layer to layer within a particular DNN model. 

(2) We present a new inter-layer dataflow mechanism that enables compression format conversions without explicit hardware modules. 

(3) We design \AcceleratorName, which hinges on a novel network topology (called MRN) that allows, for the first time, support for the three dataflows, and a new L1 on-chip memory organization to effectively  capture the memory access patterns that each dataflow exhibits for input, output and partial sums. 

(4) We extensively evaluate \AcceleratorName~using cycle-level simulations of several contemporary DNN models from different application domains, and RTL implementation of its principal elements. Our results demonstrate that Flexagon achieves average performance benefits of 4.59$\times$ (ranges between 2.09$\times$ and 7.41$\times$), 1.71$\times$ (ranges between 1.04$\times$ and 4.87$\times$), and 1.35$\times$ (ranges between 1$\times$ and 2.13$\times$) with respect to the state-of-the-art SIGMA-like, Sparch-like and GAMMA-like accelerators (265\% , 67\% and 18\%, respectively, in terms of average performance/area efficiency).

\section{Background} 
\label{section:background}


\vspace{-2mm}
\subsection{Compression formats}
\vspace{-1mm}
Following the same taxonomy used in ExTensor~\cite{hegde2019extensor}, the SpMSpM operation computes the operation $C_{M, N}=A_{M, K}\times B_{K, N}$, where the three matrices are 2-dimensional tensors. 
 Since these matrices are typically sparse (see Table~\ref{table:dnn_models}), they are compressed 
to encode the non-zero values while preserving the computation intact (lossless compression)~\cite{taco2017}.  In our work, we focus on the widely used unstructured compression formats CSR and CSC. 
A matrix encoded in CSR format employs three \mbox{1-dimensional} tensors to store the non-zero values in a row-major data layout: a  data vector to represent the non-zero values, a row pointer vector to store the index position where each row begins within the data vector, and a column index vector to store the column of each non-zero value. Similarly, the CSC uses column-major data layout: a data vector, a column pointer vector to store the index position of start of column, and a row index vector to store the row index of each non-zero data value. Observe that both CSR and CSC employ the same compression method, and thus, can be seen as a single compression format. This is important as an accelerator would use the same control logic needed to handle both of them. This facilitates the implementation of the control logic (further details in Section~\ref{section:memory_controllers}) in our accelerator.

As in previous works (e.g.~\cite{zhang2021gamma}), we will use the term \textit{fiber} to denote each compressed row or column. Each fiber contains a list of duples (coordinate, value), sorted by coordinate. We use the term \textit{element} to refer to one duple in the fiber. 

\subsection{SpMSpM dataflows}
\vspace{-1mm}
SpMSpM operation is based on a triple-nested for-loop that iterates over \texttt{A}'s and \texttt{B}'s independent dimensions \textit{M} and \textit{N}, and co-iterates over their shared dimension \textit{K}. Depending upon the level of the co-iteration in the loop nesting, \textit{three different dataflows} have been identified for SpMSpM computation: \innerprod~(co-iteration at the innermost loop), \outerprod~(co-iteration at the outermost loop) and \gustavsons~(co-iteration at the middle loop). 
Additionally, these dataflows result in \textit{six possible variants} according to how the independent dimensions (\textit{M} and \textit{N}) are ordered for each of them (two variants per dataflow). Notice that each variant favors the stationarity of one of the dimensions (the outermost one) over the other. This way, we distinguish each variant by \texttt{(M)} if the computation is \textit{M}-stationary or \texttt{(N)} if it is \textit{N}-stationary. 
Fig.~\ref{fig:dataflows} shows the resulting six dataflow variants. Each dataflow defines how the elements flow during execution, and thus, the opportunities for data reuse. Table \ref{table:dataflows_table} gives a detailed taxonomy of each approach, which we summarize as follows:


 \begin{figure}[t]
        \begin{center}
                \includegraphics[width=0.9\columnwidth, height=\columnwidth]{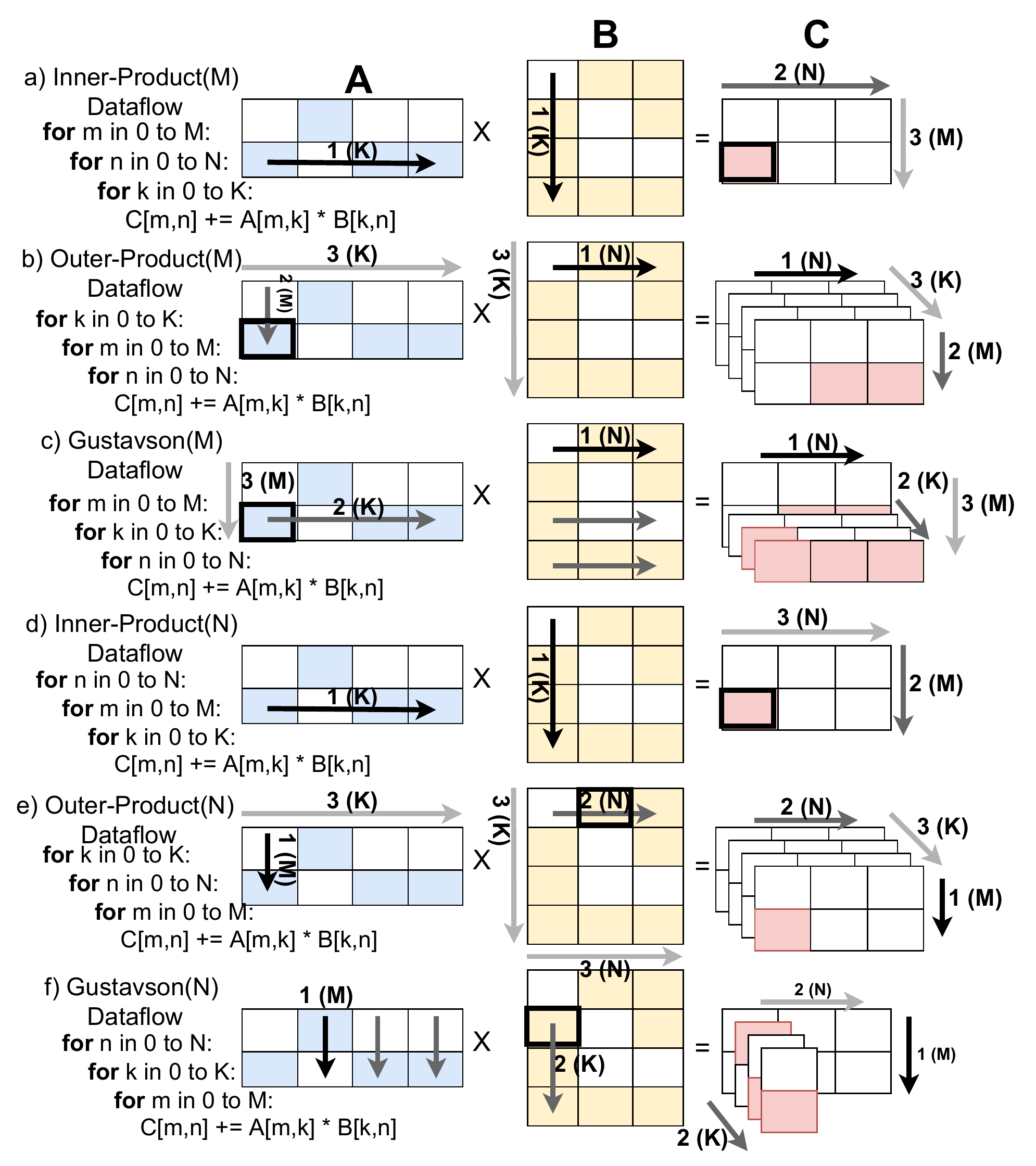}
        \end{center}
        \vspace{-0.40cm}
        \caption{Dataflow combinations for matrix multiplication. For simplicity, non-compressed (dense) matrices are shown.}
        \label{fig:dataflows}
\end{figure}

\begin{table*}[t!]
\begin{scriptsize}
\begin{center}
{
\begin{tabular}{|r|c|c|c|c|c|c|c|c|c| }\hline
\textbf{Dataflow} & \textbf{Informal Name} & \textbf{Stationary} & \textbf{Stationary} & \textbf{Streaming}
 & \textbf{A format} & \textbf{B format} & \textbf{C format} & \textbf{Intersection} & \textbf{Merging} \\
  &   &  \textbf{Tensor} &  \textbf{Fiber} &  \textbf{Tensor} &  &  &  & & \\\hline
 \textbf{MNK} & Inner Product(M) & C & A & B & CSR  & CSC & CSR & Scalar A vs Scalar B & N/A \\\hline
 \textbf{KMN} & Outer Product(M) & A & B & C & CSC & CSR & CSR  & N/A & Scalar \\\hline
 \textbf{MKN} & Gustavson's(M) & A & C & B & CSR  & CSR & CSR  & Scalar A vs Fiber B & Fiber(M)\\\hline
 \textbf{NMK} & Inner Product(N) & C & B & A & CSR  & CSC & CSC  & Scalar B vs Scalar A & N/A \\\hline
 \textbf{KNM} & Outer Product(N) & B & A & C & CSC  & CSR & CSC  & N/A & Scalar \\\hline
 \textbf{NKM} & Gustavson's(N) & B & C & A & CSC  & CSC & CSC  & Scalar B vs Fiber A & Fiber(N)\\\hline
\end{tabular}
}
\end{center}
\end{scriptsize}
\vspace{-0.35cm}
\caption{Taxonomy of dataflow properties. Traversal order is given outermost-to-innermost in loop order. }
\label{table:dataflows_table}
\vspace{-0.25cm}
\end{table*}

\textbf{Inner Product (IP)}: A single full sum at a time is generated, with no merging of partial sums. This requires a hardware intersection unit to align effectual inputs.

\textbf{Outer Product (OP)}: A single input scalar at a time is consumed, generating many partial sums. This does not require intersection but does require merging hardware, and possibly extra memory traffic.

\textbf{Gustavson's (Gust)}: A single input at a time is consumed, but only to generate partial sums into the current fiber. This allows the intersection to be done in \emph{leader-follower} style, where the effectual coordinates of the stationary tensor retrieve entire fibers of the other operand. This requires merge hardware, but only into the current fiber rather than entire matrices.

For the rest of the paper,  we will pedagogically use \mbox{\textit{M}-stationary} dataflows during the explanations, although everything would apply for the \textit{N}-stationary dataflows as well.


 \begin{figure}[!t]
        \begin{center}
                \includegraphics[width=0.8\columnwidth, height =0.7\columnwidth]{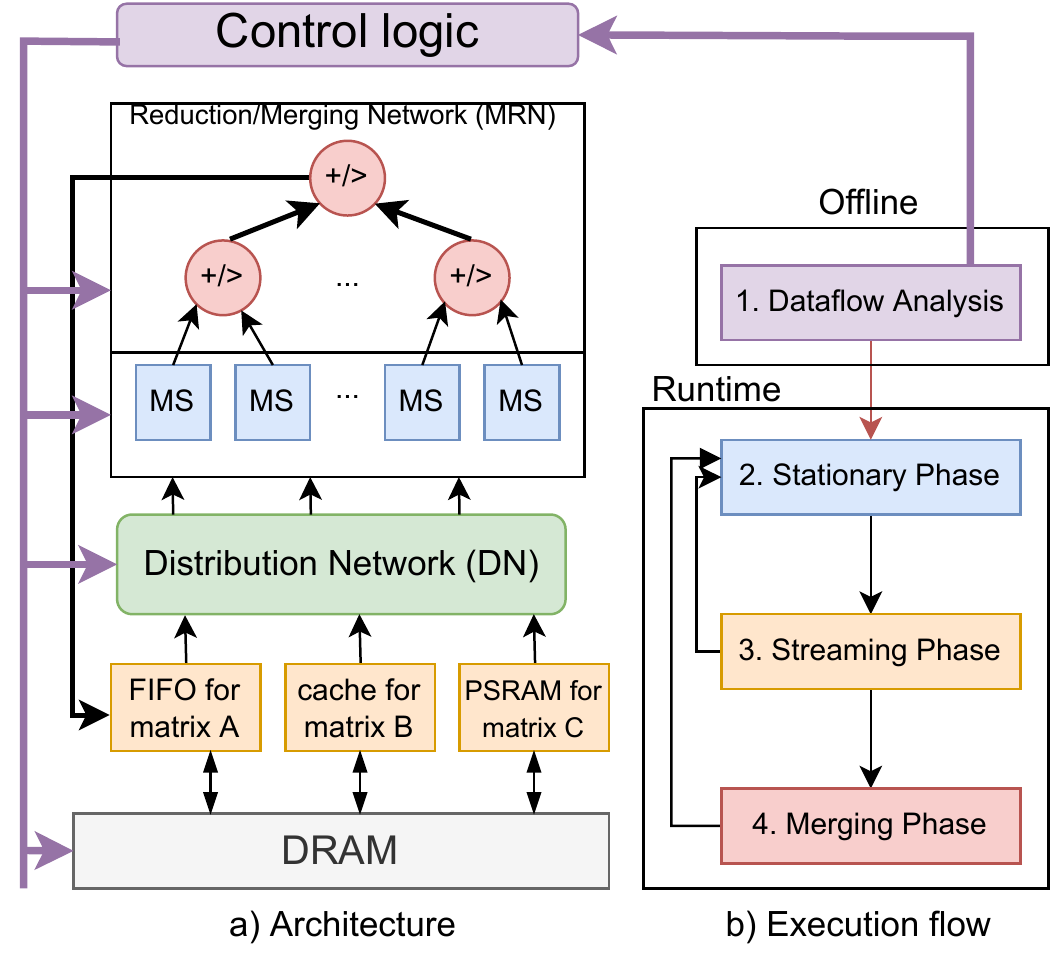}
        \end{center}
       \vspace{-0.40cm}
        \caption{\AcceleratorName~high-level overview.}
        \label{fig:\AcceleratorName_overview}
        \vspace{-0.1cm}
\end{figure}

\vspace{-1mm}
\section{{\AcceleratorName}~Design}  
\label{section:design}
\vspace{-1mm}
Fig.~\ref{fig:\AcceleratorName_overview}a shows a high-level overview of the architecture of the \AcceleratorName~accelerator. As observed,  \AcceleratorName~consists of a set of multipliers, adders and comparators, as well as three on-chip SRAM modules specifically tailored to the storage needs of matrices \texttt{A}, \texttt{B} and \texttt{C} for the three SpMSpM dataflows. In addition, in order to allow for the highest flexibility,  all the on-chip components are interconnected by using a general three-tier reconfigurable network-on-chip (NoC) composed of a Distribution Network (DN), a Multiplier Network (MN), and a Merger-Reduction Network (MRN), inspired by the taxonomy of on-chip communication flows within
AI accelerators~\cite{MAERI2018}.  These components are controlled by the control unit which is configured by the mapper/compiler before the execution.  

\AcceleratorName's execution phases are shown in Fig.~\ref{fig:\AcceleratorName_overview}b. The process begins with a dataflow analysis (phase 1), which is carried out offline. Here, a mapper/compiler examines the features of the SpMSpM operation to be executed (i.e., matrix dimensions and sparsity patterns) and decides the dataflow (between the six available described in Section~\ref{section:background}) that best matches the operation, generating the tiling scheme and the particular values for the signals that configure the operation of the accelerator for the rest of the phases.

The next three phases are performed during runtime according to these generated signals and are repeated several times according to the number of execution tiles. The first runtime phase is called \textbf{stationary phase} (phase 2), which delivers data that will be kept stationary in the multipliers to reduce the numer of costly memory accesses. According to the dataflows description presented in Section~\ref{section:background} for \mbox{\textit{M}-stationary} dataflows, this stationary data belongs to matrix \texttt{A}, while matrix \texttt{B} is streamed during the \textbf{streaming phase} (phase 3). For \textit{N}-stationary dataflows this happens in the reverse order. These two phases generalize for the three dataflows. The \textbf{merging phase} (phase 4) is only necessary for both \outerprod~and \gustavsons~dataflows and is the one in charge of merging the fibers of partial sums that have been previously generated during the streaming phase. This phase is skipped in the \innerprod~dataflow as no merging is required. 

In this work, we focus our attention on the accelerator design as well as on the way the three phases operate in order to give support to the six possible dataflows (three SpMSpM dataflows, two variants, M or N-stationary, each) over the same hardware substrate. We leave the study of the tool required for dataflow analysis, tiling selection and generation of the configuration file for the accelerator (phase 1 in the Offline part in Fig.~\ref{fig:\AcceleratorName_overview}b) for future work.

\subsection{On-chip Networks}
\vspace{-1mm}
One of the main novelties of \AcceleratorName~is its ability (through proper configuration) to support the six dataflows described in Section~\ref{section:background} using the same hardware substrate. To do so, the accelerator is equipped with a three-tier configurable NoC able to adapt to the communication features of each dataflow. Next, we describe each subnetwork in detail:

\textbf{Distribution network} (DN): This module is used to deliver data from the SRAM structures to the multipliers. In order to enable the high flexibility that the three SpMSpM dataflows require, the DN needs to support unicast, multicast and broadcast data delivery. To achieve this, and at the same time ensure high energy efficiency, we utilize a Benes network similar to previous designs like SIGMA~\cite{SIGMA2020}. This network is an N-input, N-output non-blocking topology with $2~\times~log(N)+1$ levels, each with N tiny 2$\times$2 switches.

\begin{figure*}[t!]
        \begin{center}
                \includegraphics[width=1.8\columnwidth]{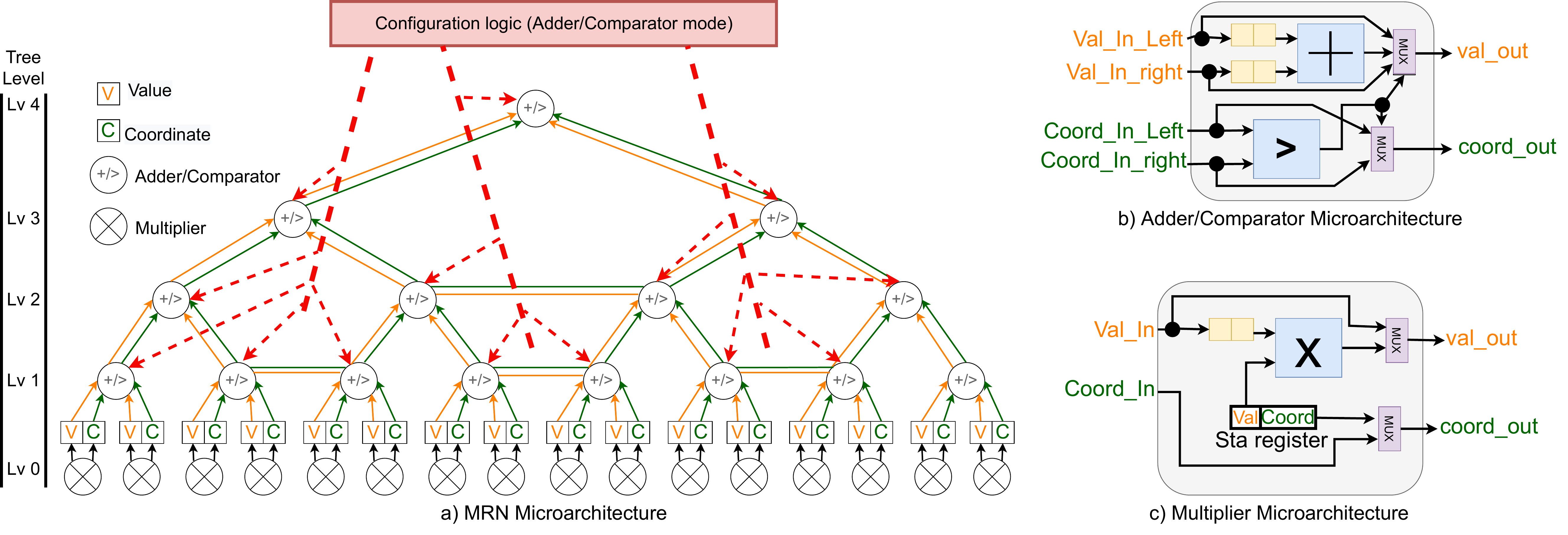}
        \end{center}
        \vspace{-0.40cm}
        \caption{a) MRN topology. b) $\micro$architecture of the MRN's nodes (Adder/Comparator nodes). c) $\micro$architecture of Multipliers.
        } 
        \label{fig:unified_rn}
\end{figure*}

\textbf{Merger-Reduction network} (MRN): Previous designs like MAERI~\cite{MAERI2018} or SIGMA~\cite{SIGMA2020} have used specialized tree-based reduction networks (RNs) such as ART or FAN to enable non-blocking reduction of multiple clusters of psums. These RNs provide high flexibility for the \innerprod~dataflow as its purpose is to reduce a cluster of psums. In case of \outerprod~and \gustavsons~dataflows, other works such as~\cite{zhang2021gamma, SpArch2020} employ a tree-based topology to perform the merge operation of the psums once they are generated. In our design, we have, for the first time, unified this concept, and have designed a merger-reduction network able to both reduce and merge psums. Figure~\ref{fig:unified_rn}a shows the microarchitecture overview of a 16-wide MRN. As it may be observed, similar to previous designs, we also employ an augmented tree-based topology because this is the fastest way to perform both the reduction and merging operations. Differently to previous designs, the MRN topology augments the nodes with comparators and switching logic able to exchange the mode of operation (see Figure~\ref{fig:unified_rn}b). This allows to perform both operations while keeping low area and power overheads (details in Section~\ref{section:rtl_results}) and, as we describe later, enables direct support for the three SpMSpM dataflows. Furthermore, we employ a connection with two links between the nodes, allowing the MRN to traverse not only data values but also the coordinates needed in both \outerprod~and \gustavsons~dataflows. 
The selection of the configuration is done by the mapper/compiler, which generates the control signals that feed the configuration logic module (Control Logic in Fig.~\ref{fig:\AcceleratorName_overview}a) of the accelerator, which in turns routes the appropriate signals to the nodes, configuring its operation modes according to the dataflow and layer dimensions.  

\textbf{Multiplier network} (MN): Similar to other designs such as MAERI, this network is composed of independent multipliers that can operate in two different modes: i) \textit{Multiplier mode}: the unit performs a multiplication and sends the result to the MRN. This mode is used during the entire execution when the \innerprod~dataflow is configured, and during the streaming phase when either the \outerprod~or \gustavsons~dataflows are configured; ii) \textit{Forwarder mode}: the multiplier forwards directly the input, which is typically a psum, to the MRN. As we will clarify in the examples presented next, this mode is essentially configured during the merging phase in both the \outerprod~and \gustavsons~dataflows. 
The microarchitecture of the multipliers is depicted in Figure~\ref{fig:unified_rn}c.

\subsection{Walk-through Examples}

Next, we illustrate how \AcceleratorName~works when running the three dataflows for the multiplication of matrices \texttt{A} and \texttt{B} from Fig.~\ref{fig:dataflows}, considering the runtime phases explained earlier.
We pedagogically assume the \innerprodM, \outerprodM~and \gustavsonsM~dataflows. Note that, the \innerprodN, \outerprodN~and \gustavsonsN~dataflows could be executed in the same manner by exchanging matrices \texttt{A} and \texttt{B}. To ease the explanation, we assume a simple 4-multiplier accelerator, and we walk through the activity of the three sub-networks. In the explanation, we mention the on-chip SRAM modules needed for storing matrices \texttt{A}, \texttt{B}, \texttt{C} and psums (see the yellow boxes in Fig.~\ref{fig:\AcceleratorName_overview}b). Section~\ref{section:memory_organization} provides an in-depth description 
of these 
memory structures.

\subsubsection{Example of Inner-Product dataflow.}
\label{section:example_inner_product}
\begin{figure}[t!]
        \begin{center}
                \includegraphics[width=1.0\columnwidth]{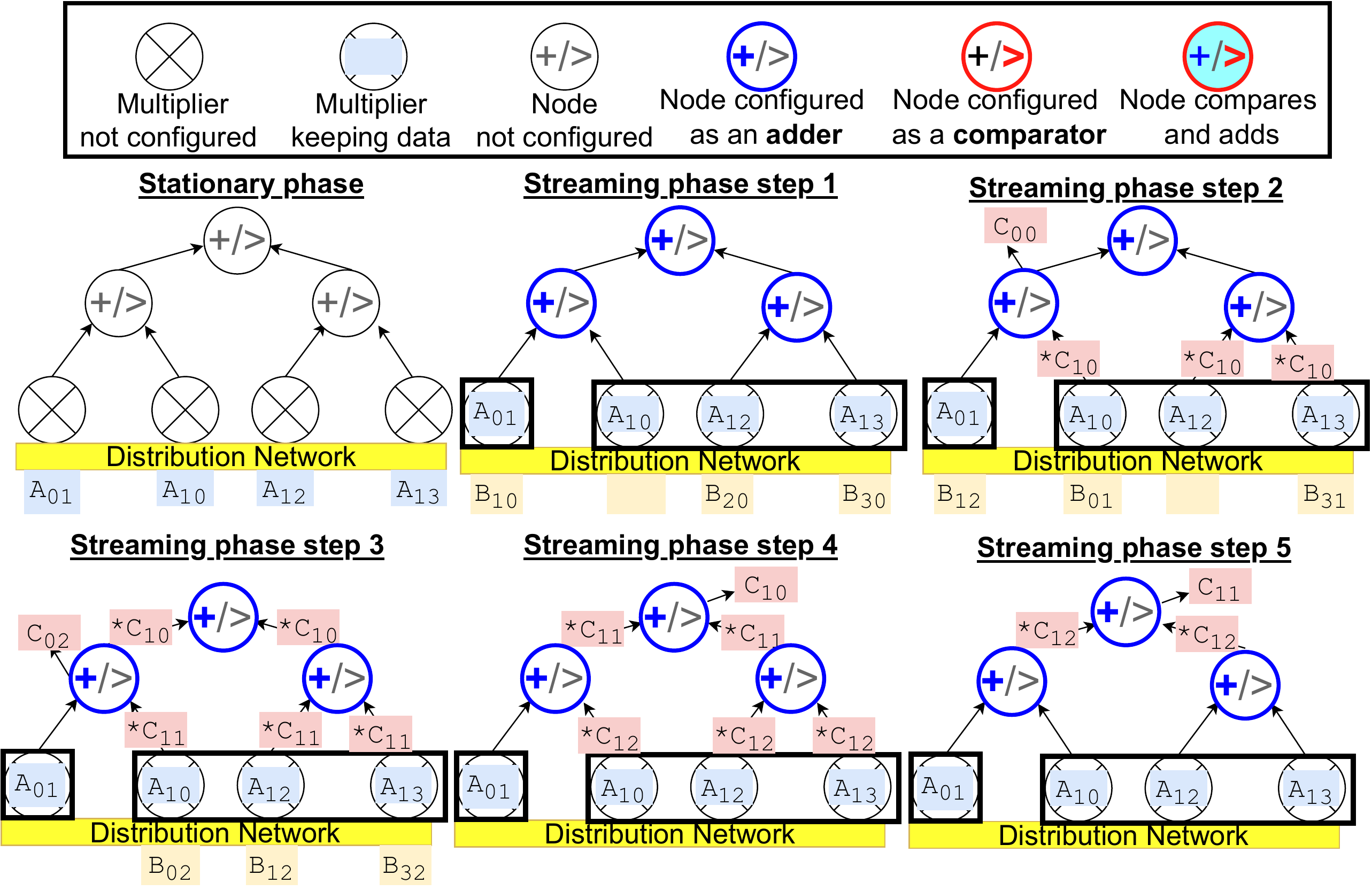}
        \end{center}
        \vspace{-0.40cm}
        \caption{Example of \AcceleratorName~running SpMSpM using an \texttt{Inner-Product(M)} dataflow. ``\textit{*}'' indicates psums.}
        \label{fig:inner_product_example}
        \vspace{-0.15cm}
\end{figure}

Fig.~\ref{fig:inner_product_example} shows the \innerprodM~dataflow. In the figure, we represent with ``\textit{*}'' the psums that need to be reduced by the adders in the tree to produce the final values for matrix \texttt{C}.

\textbf{Stationary phase}: First, during the stationary phase, the controller maps as many fibers of matrix \texttt{A} (i.e., rows of \texttt{A}) as possible to the multipliers, reading all the elements  sequentially from the dedicated SRAM structure called FIFO for matrix \texttt{A}. Each cluster of multipliers will perform the dot product operation. 

\textbf{Streaming phase}: After filling the multipliers with the fibers of \texttt{A}, the controller multicasts each fiber of matrix \texttt{B} (i.e., each column) to the configured clusters in the MN. To do so, the controller uses the row coordinate of each element in the fiber of \texttt{B} to detect whether it intersects with the column coordinate in the fiber of \texttt{A}. If this happens, the value is sent out to the corresponding multiplier.


\subsubsection{Example of Outer-Product dataflow.}
\label{section:example_outer_product}
\begin{figure*}[t!]
        \begin{center}
                \includegraphics[width=1.3\columnwidth]{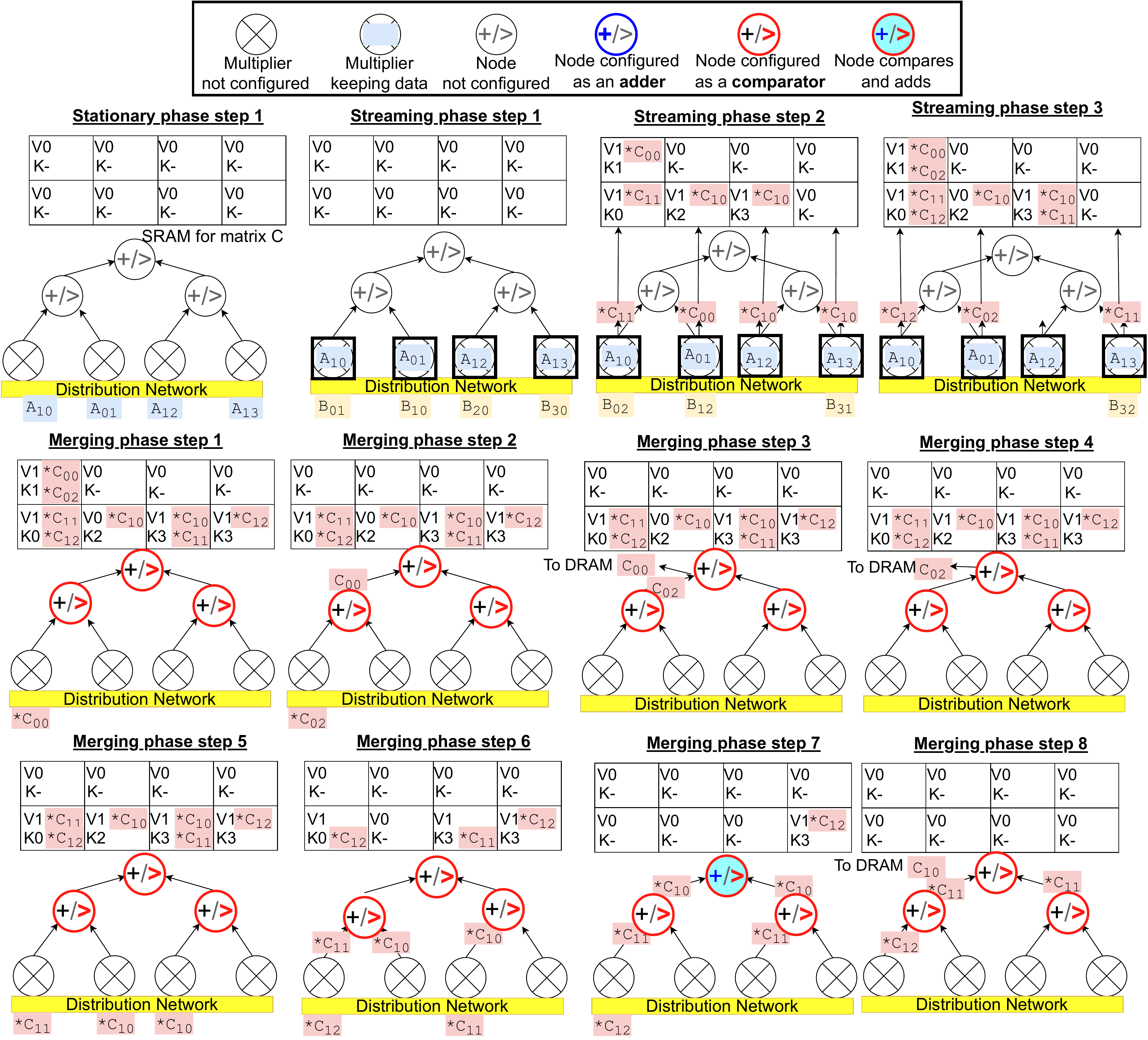}
        \end{center}
        \vspace{-0.30cm}
        \caption{Example of \AcceleratorName~running SpMSpM using an \texttt{Outer-Product(M)} dataflow. ``\textit{*}'' indicates that the outputs produced by the accelerator are psums and not final outputs. ``\textit{V}'' in the \SramCName represents the valid bit and ``\textit{K}'' indicates the \textit{k} iteration tagged for a particular block.}
        \label{fig:outer_product_example}
\end{figure*}

Fig.~\ref{fig:outer_product_example} shows the same example as before but now assuming the \outerprodM~dataflow. We also show the customized SRAM structure for \texttt{C} called \SramCName that is utilized for storing the psums for matrix \texttt{C}. As we will explain in Section~\ref{section:memory_organization}, this structure stores blocks of elements (coordinate, value).

\textbf{Stationary phase}: During the stationary phase, the fibers of matrix \texttt{A} (i.e., columns of \texttt{A}) are delivered to the multipliers sequentially following the CSC compression format. In our particular case, the four multipliers store the elements $A_{1,0}$, $A_{0,1}$, $A_{1,2}$, and $A_{1,3}$.

\textbf{Streaming phase}: During the streaming phase, each multiplier keeps stationary an element  $A_{m,k}$, given \textit{m} in range [0,\textit{M}) and \textit{k} in range [0,\textit{K}), in order to linearly combine the non-zero elements $B_{k,0:N-1}$, generating a psum fiber where all the elements share the row (\textit{m}) and a particular \textit{k} iteration (i.e., the partial matrix where these elements belong to). Consecutive multipliers generating psums for different rows for the same \textit{k} iteration, do not need the psum to be merged together. Thus, the generated psums must be sent out to the SRAM structure, in order to be merged in a third phase. Also, since multiple rows can run in parallel, the \SramCName's set is indexed by rows. Furthermore, since the number of non-zeros in matrix \texttt{A} is not known a priori, it might happen that multiple fibers from matrix \texttt{A} may fit in a single iteration, causing that multiple partial outputs for the same row, but for different \textit{k} iterations, may run in parallel.
Since the number of psums for a particular row  and for a particular \textit{k} iteration is not known at runtime, we must assign static space in the \SramCName to store the psums from different \textit{k} iterations that may be running in parallel and being kept in the same row. To do so, we divide each row in the \SramCName in blocks, and each block contains a valid bit to indicate the validity of the data, a \textit{k} value, indicating the \textit{k} iteration that belongs to that group of partial sums and the block of data. By doing this, each block can hold, at a particular time, psums for different \textit{k} iterations for a particular row. This way, if the number of psums for a particular iteration exceeds the block size, it may use another block from the row, even if the next block is already being used by another \textit{k} iteration.
The details about the organization and operation of the \SramCName are given in Section~\ref{section:memory_organization}.

In the example of Fig.~\ref{fig:outer_product_example}, we see three steps regarding the streaming phase. In the first step, the controller sends the first element of the four fibers (across the \textit{K}-dimension) to its corresponding multiplier. For example, the first multiplier which keeps stationary the element $A_{1,0}$ receives the first element of the fiber for the row (i.e., iteration \textit{k}) 0. In step 2, each multiplier generates a psum (indicated by the symbol \textit{*}), which is the first element for the 4 fibers generated across the \textit{K}-dimension. These psums are then stored in the \SramCName. The first psum *$C_{1,1}$ is allocated in set 1, as it is indexed by its row coordinate. Use of sets allows us to execute multiple rows in parallel. Then, since the first line is free, the psum is stored there, enabling the valid bit and indicating that the element belongs to \textit{K0}. Dividing rows into blocks allows holding psums corresponding to different \textit{K} for a particular row. 
The second psum, *$C_{0,0}$ is allocated to the set 0 (its row coordinate) and since the first line is here, the cache enables the valid bit and tags the line with \textit{K1}. The last two elements share coordinates (i.e., s *$C_{1,0}$), but belong to a different partial matrix (\textit{K2} and \textit{K3}). These two elements go to the same set in the \SramCName  but to different lines, each tagged with its iteration \textit{k} (i.e., \textit{K2} and \textit{K3}). This allows to locate the psum fibers in the correct order during the merging phase.

In step 3, the second elements for the four fibers are produced, following the same execution scheme. For the sake of brevity, we do not show how the last element from the longest psum fiber (i.e., fiber K3) is produced, and directly show the contents of the \SramCName just before starting the merging phase (merging phase step 1). We can see in the \SramCName  figure from the merging phase step 1 that the element has been stored in the last line within the first set, as the third line is already full. 

\textbf{Merging phase}: The merging phase proceeds row by row. For each row, the controller fetches the elements for the different \textit{k}-iteration fibers from the \SramCName. These elements are stored in different blocks and can be identified by their tags, consuming the elements and sending them to the MRN in order to be merged. Each  unit in the MRN compares the column coordinate (i.e., the \textit{N}-dimension). If the coordinates match, then the values of the elements are accumulated. Otherwise, the node sends up the tree the element with the lowest coordinate.  The last two rows in Fig.~\ref{fig:outer_product_example} show 8 merging steps. The 4 first steps (Merging phase step 1 to step 4) merge the first row. In the second row, there are 3 psum fibers ready to be merged. In step 5, the first elements for the three fibers (\textit{K0}, \textit{K2} and \textit{K3}) are sent to the MRN. In step 6, the psums *$C_{1,1}$ and *$C_{1,0}$ compare their column coordinate. Since they do not match, and element *$C_{1,0}$ has a lowest column coordinate, this element is sent up to the MRN first.  The same procedure is executed in a pipelined-manner for the rest of the elements in the fiber until all the psums have been merged in a single fiber and sent to DRAM. In case the number of fibers in a row is greater than the number of multipliers (i.e., leaves in the tree), the controller needs to perform multiple passes to complete the final merge.

 \begin{figure}[t!]
        \begin{center}
                \includegraphics[width=1.0\columnwidth]{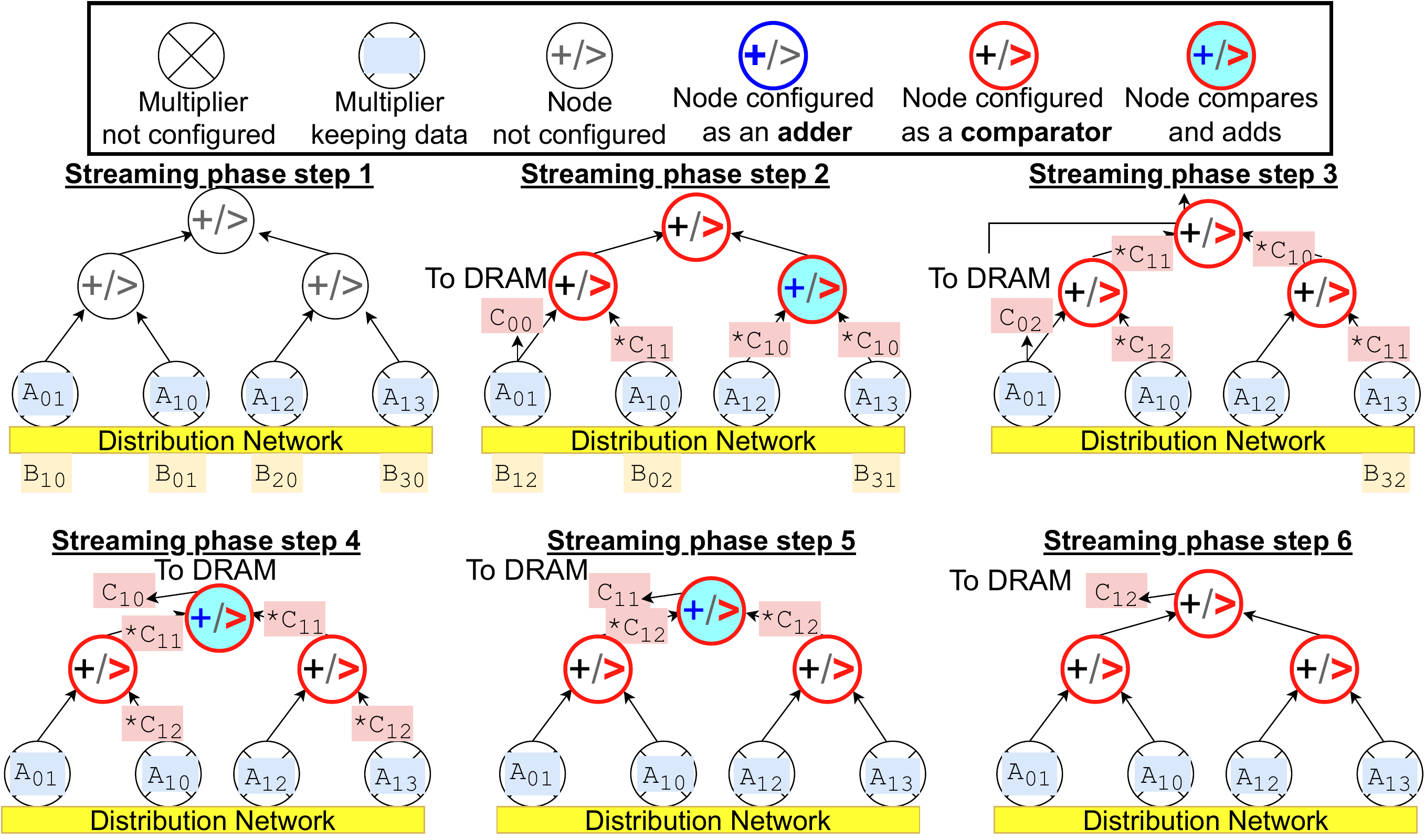}
        \end{center}
        \vspace{-0.30cm}
        \caption{Example of \AcceleratorName~running SpMSpM using the \texttt{Gustavson(M)} dataflow. ``\textit{*}'' indicates that the outputs produced by the accelerator are psums and not final outputs.}
        \label{fig:gustavsons_product_example}
        \vspace{-0.15cm}
\end{figure}

\subsubsection{Example of Gustavson's dataflow.}
\label{section:example_gustavsons}
Finally, for the same example matrices, Fig.~\ref{fig:gustavsons_product_example} illustrates how \AcceleratorName~proceeds when the \gustavsonsM~dataflow is selected. Similarly, the operation in this case proceeds in three well-differentiated phases.

\textbf{Stationary phase}: First, during the stationary phase, as many fibers of \texttt{A} (i.e., rows in matrix \texttt{A}) as possible are mapped spatially and sequentially in the multipliers. The multipliers, then keep two clusters, each in charge of calculating the psums for a different output row (i.e., rows 0 and 1 in the example).

\textbf{Streaming phase}: In the streaming phase, for each multiplier, the memory controller fetches and delivers the fiber of \texttt{B} (i.e., row of \texttt{B}) that corresponds to the column coordinate (i.e., \textit{k}-iteration) associated to the mapped element of \texttt{A} in the multiplier.  Every multiplier generates a partial output fiber which is merged with the rest of partial output fibers generated by the other multipliers allocated to the same fiber of \texttt{A}. An example of this generation is shown in Fig.~\ref{fig:gustavsons_product_example}. Here, we depict 6 streaming steps. The first multiplier keeps stationary the only one  element in matrix \texttt{A} ($A_{0,1}$) so it receives the fiber of \texttt{B} indexed by column 1 (i.e., row 1). The second, third and fourth multipliers keep the elements $A_{1,0}$, $A_{1,2}$ and $A_{1,3}$, respectively, so they receive the fibers of \texttt{B} 0, 2 and 3, respectively. The first 3 steps show how the elements from the fibers of \texttt{B} are delivered cycle by cycle. 

\textbf{Merging phase}: Similar to the \outerprod~dataflow, the merging phase combines both the accumulation and the merging operation, accumulating the elements (i.e., its values) in a certain node if their column coordinates match, or sending the element with the lowest column coordinate value. On the other hand, in \gustavsons~dataflow, we can merge the psums immediately after their generation, as a cluster of multipliers always generates fibers for the same row, but for different \textit{k} iterations. When the number of elements in \texttt{A} fits into a cluster of multipliers, the output fiber generated by that cluster will be a final fiber, and the outputs can be sent directly to DRAM without being stored in the SRAM. Otherwise, when the number of elements in \texttt{A} exceeds the number of multipliers, the output fiber will be a partial fiber as multiple iterations are required, and therefore the fiber will require to be stored in the PSRAM, similar to what happens in the \outerprod~dataflow. 

\begin{figure}[t!]
        \begin{center}
                \includegraphics[width=0.9\columnwidth]{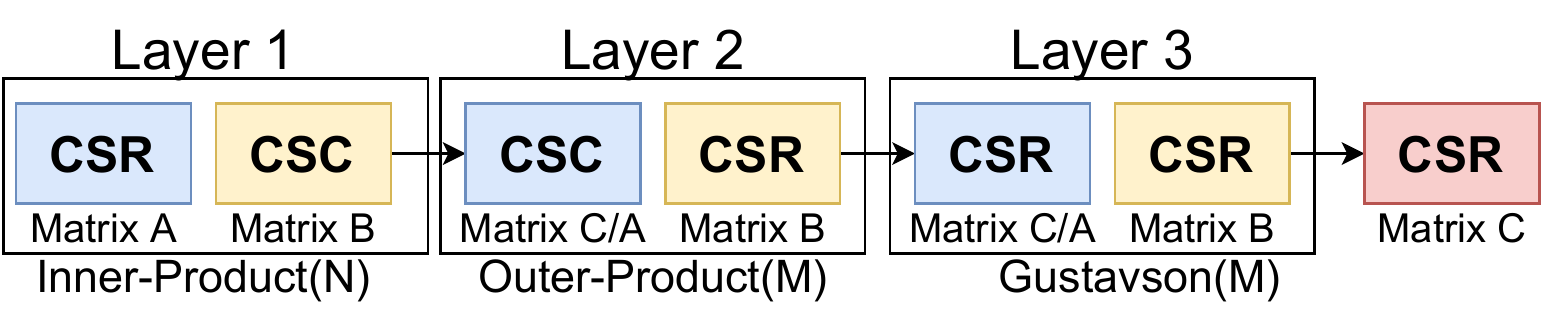}
        \end{center}
        \vspace{-0.40cm}
        \caption{Example of three DNN layers being executed by running the combination \texttt{Inner-Product}, \texttt{Outer-Product} and \texttt{Gustavson} dataflows.}
        \label{fig:dataflow_combinations}
    \vspace{-0.15cm}
\end{figure}

\subsection{Combinations of inter-layer dataflows}

As Table~\ref{table:dataflows_table} shows, \textit{M}-stationary dataflows output the elements in CSR format while \textit{N}-stationary dataflows output the elements in CSC format. \AcceleratorName~supports the six dataflows and takes advantage of this observation to appropriately execute every possible sequence of DNN layers without requiring costly explicit hardware format conversions and is the first work to support compressed outputs without explicit conversions. Fig.~\ref{fig:dataflow_combinations} shows an example of a DNN composed of three layers, demonstrating this feature. The first and the second layer are configured to execute inner and outer products respectively. Since second layer needs activation in CSC, first layer is Inner Product (N). The weights are assumed to be stored offline in both formats. The second layer produces the matrix in CSR format if it uses M-stationary. As a result it could choose from inner product or Gustavsons(M).

Table~\ref{table:dataflows_transitions} shows the transitions for each dataflow combination that do not require an explicit format conversion (green tick) and those that do (\textit{Explicit Conversion} or EC). 
These combinations can utilized by the mapper/compiler to generate the best sequence of dataflows that lead to the best performance and energy efficiency for a particular DNN execution.

\begin{table}[t!]
\begin{footnotesize}
\begin{center}
{
\begin{tabular}{|c|c|c|c|c|c|c|}\hline
 & \textit{IP(M)} & \textit{OP(M)} & \textit{Gust(M)} & \textit{IP(N)} & \textit{OP(N)} & \textit{Gust(N)} \\\hline
 \textit{IP(M)} & \greencheck  & EC & \greencheck  & \greencheck & EC & EC\\\hline
 \textit{OP(M)} & \greencheck  & EC & \greencheck  & \greencheck & EC & EC\\\hline
 \textit{Gust(M)} & \greencheck  & EC & \greencheck  & \greencheck & EC & EC\\\hline
 \textit{IP(N)} & EC  & \greencheck & EC  & EC & \greencheck & \greencheck\\\hline
 \textit{OP(N)} & EC  & \greencheck & EC  & EC & \greencheck & \greencheck\\\hline
 \textit{Gust(N)} & EC  & \greencheck & EC  & EC & \greencheck & \greencheck\\\hline

\end{tabular}
}
\end{center}
\end{footnotesize}
\vspace{-0.35cm}
\caption{Dataflow transitions allowed without requiring Explicit format Conversion (EC). Different rows represent the different outputs of the first layer and different columns represent the corresponding input to the second layer.  
}
\label{table:dataflows_transitions}
\vspace{-0.2cm}
\end{table}

\subsection{Memory organization}
\label{section:memory_organization}
\begin{figure}[t!]
        \begin{center}
                \includegraphics[width=1.0\columnwidth]{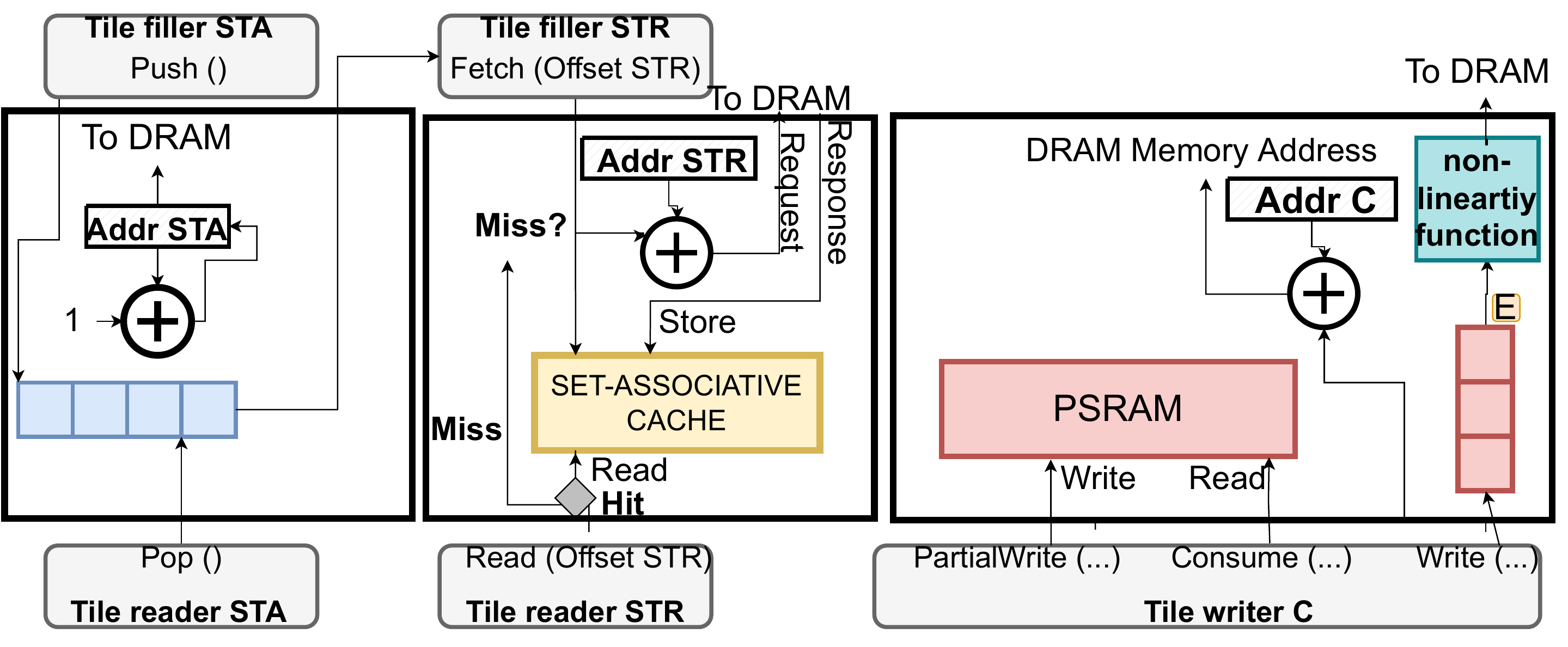}
        \end{center}
        \vspace{-0.40cm}
        \caption{Memory structures in \AcceleratorName.}
        \label{fig:l1_memory_organization}
\end{figure}

\vspace{-1mm}
In order to capture all dataflows, we have designed a customized L1 memory level specifically tailored for the common and different patterns among the three dataflows. 
Fig.~\ref{fig:l1_memory_organization} shows a schematic design for this L1 memory level. We use a separate memory structure and a different buffer idiom for data movement from/to each structure. 
To do so, every memory structure is operated by two controllers, the \textbf{tile filler}  interfacing with the DRAM, and the \textbf{tile reader} interfacing the datapath of the accelerator (i.e., the multipliers). Next, we describe each memory structure in detail: 

\textbf{Memory structure for the stationary matrix (FIFO)}: The elements of the stationary matrix are always read once and sequentially for the three dataflows, as they are kept stationary in the multipliers. To hide the access latency, we implement a 512-byte read-only FIFO.
In order to save bandwidth and reduce the complexity: (1) the memory structure keeps the DRAM location of the stationary matrix in a register, so that the fibres are pushed implicitly into FIFO, (2) we employ a single-port for read and write.

\textbf{Memory structure for the streaming matrix (Cache)}: The streaming matrix presents a more heterogeneous memory access pattern. 
In \innerprod, every stationary-phase (i.e., every iteration) causes the  streaming of the entire matrix. In other words,  there is significant spatial locality and temporal locality every time the matrix is re-loaded.  In the \outerprod~dataflow, the fibers of the streaming matrix are read just once and sequentially. In \gustavsons~dataflow, every fiber of the stationary matrix gathers \textit{F} fibers of the streaming matrix, \textit{F} being the number of non-zero elements in the fiber of the stationary matrix which are typically scattered all over the matrix, causing an irregular and unpredictable memory access pattern. To factor the worst-case \gustavsons~dataflow, we implement the memory structure for the streaming matrix as a traditional read-only set-associative cache. However, we implement this cache to operate on a virtual address space relative to the beginning of the streaming matrix, which let us use shorter memory addresses 
and therefore save bandwidth and reduce the tag lengths.

\begin{figure}[t!]
        \begin{center}
                \includegraphics[width=0.7\columnwidth]{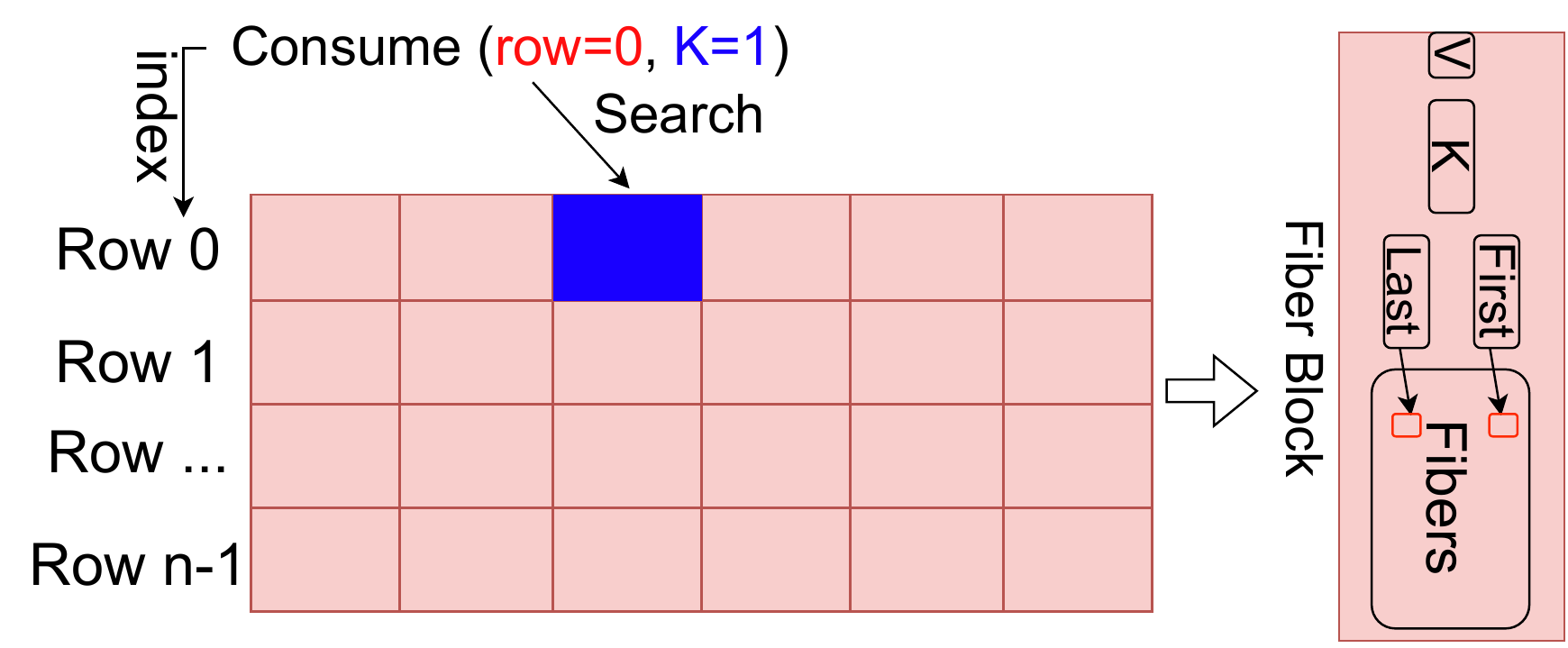}
        \end{center}
        \vspace{-0.40cm}
        \caption{PSRAM overview.}
        \label{fig:psum_cache}
\end{figure}

\textbf{Memory structure for matrix C (PSRAM)}:  To store the psums, we have designed a new buffer idiom called \textit{\SramCName}, which is used for both \outerprod~and \gustavsons~dataflows. Fig.~\ref{fig:outer_product_example} shows the way this memory structure works, Fig.~\ref{fig:l1_memory_organization} shows a high-level diagram, and Fig.~\ref{fig:psum_cache} delves into detail. The memory is organized into sets corresponding to different rows and each set into blocks for different K dimension within a row. Each block has a valid bit. 
Besides, we use a register as a line tag to keep the column coordinate (i.e., the \textit{k}-iteration) assigned to that line. Since the length of the output fiber is undetermined, it may occupy several (and non-consecutive) lines in the same row. This is essentially a way-combining scheme tagged by the \textit{k}-iteration. The register is used by the row to locate whether a certain output fiber is still placed in the PSRAM. In order to read several fibers in parallel from the same set (i.e., to merge a particular row or column) we implement a multi-bank scheme organized across the lines within a set.  Finally, we also include two registers 
to keep the byte location where the first and last elements are in the line. 

\textbf{PartialWrite(\textit{row}, \textit{k}, \textit{E})}: This operation is used to place an element in the PSRAM. The logic, indexes the element by the \textit{row} argument and then searches in parallel 
the line where is being stored the output fiber with the \textit{k} identifier. If the output fiber exists (i.e., the \textit{k} tags match), the \SramCName places the new element \textit{E} into the last available position (indicated by the register \textit{Last} in the metadata) of the last line. If the fiber does not exist, the logic searches the first available line and stores the element \textit{E} in the first position of the line, enabling the valid bit and updating the \textit{K}, \textit{First} and \textit{Last} registers in order to continue storing elements for the same \textit{K} identifier in future accesses. 

\textbf{Consume(\textit{Row}, \textit{k})}: The elements within a partial output fiber are placed in the \SramCName temporarily.  They are read once to feed the accelerator and are no longer used again. This allows us to incorporate the \textbf{consume} operation, which reads and erases a particular element from the memory structure. In particular, the controller merges the partial output fibers row by row. To do so, the controller needs to read as many fibers as possible for the same row and for each fiber it uses the \textbf{consume} operation indicating the \textit{row} and the fiber \textit{k} to search. If there is an active line keeping the \textit{k} fiber, the structure reads the next element from that fiber (indicated by the register \textit{First}) and consumes it by increasing by one element this register. When the \textit{First} and \textit{Last} registers store the same value, the \SramCName detects that the line has been consumed and invalidates the line by setting the valid bit to 0.

\textbf{Write(Offset, E)}: Apart from the \SramCName which is used to store partial output fibers, we also augment our memory structure with a FIFO which is used as a write buffer to hide the latency of sending out final output fibers to DRAM. This structure is employed by the \textbf{Write} operation.

 \begin{figure}[t!]
        \begin{center}
                \includegraphics[width=\columnwidth, height =\columnwidth]{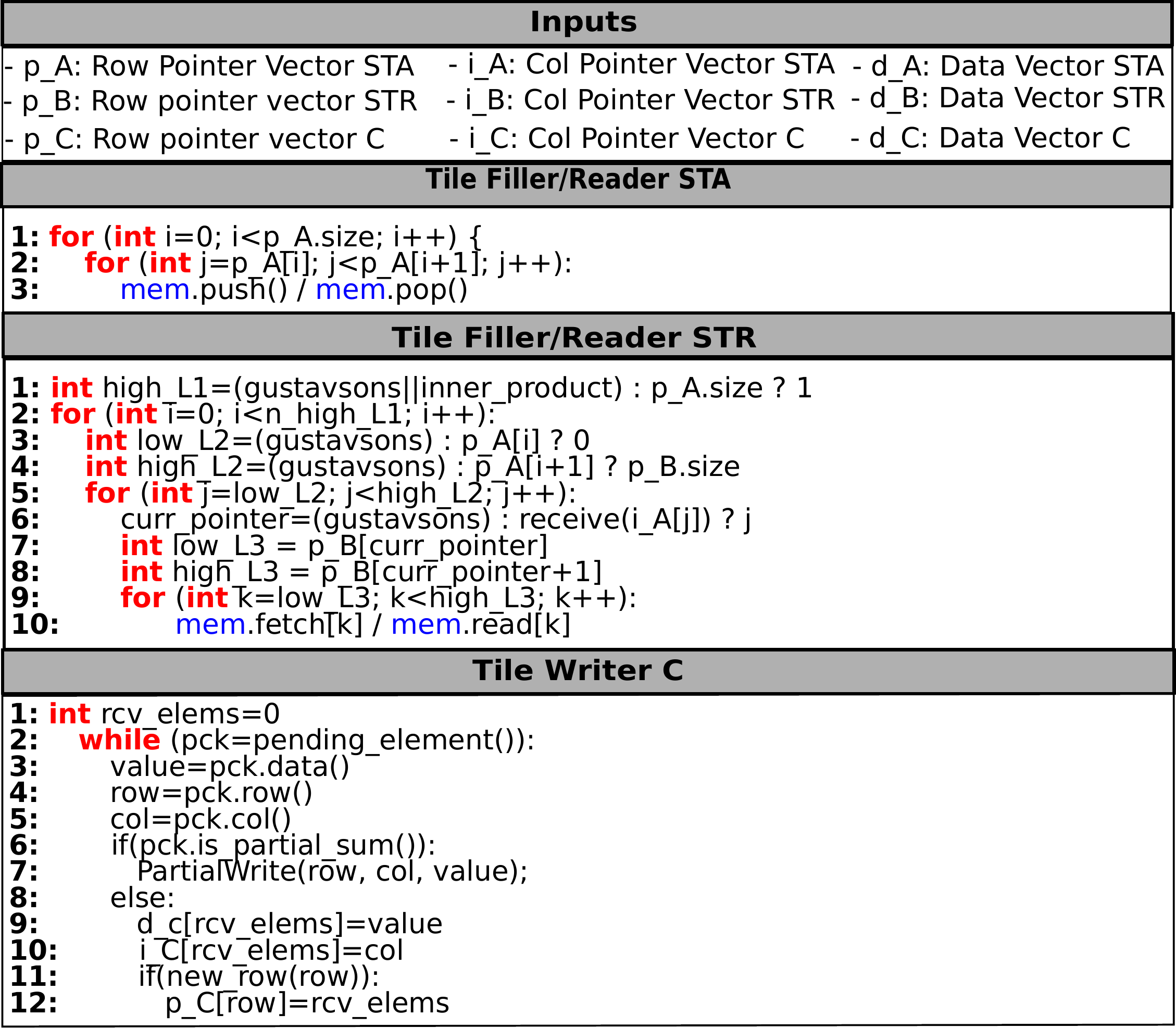}
        \end{center}
        \vspace{-0.40cm}
        \caption{Pseudo-code of the tile filler STA,  tile reader STA, tile filler STR, tile reader STR and the tile writer C. We fuse the fillers and readers in the same text box. STA: Stationary, STR: Streaming.}
        \label{fig:memory_controllers_code}
        \vspace{-0.15cm}
\end{figure}

\subsection{Memory controllers}
\label{section:memory_controllers}

Having one memory controller for each combination of dataflow and memory structure would be very costly in terms of area and power as it would require 30 logic modules to orchestrate the data (\textit{6 dataflows $\times$ 5 memory controllers}). In our design, we have unified the logic and each controller is able to be configured according to the memory access pattern of each dataflow. 
This way, as shown in Fig.~\ref{fig:l1_memory_organization}, we only need
two controllers to orchestrate the data for the memory structure which is kept stationary (i.e., the tile filler STA and the tile reader STA), two memory controllers to orchestrate the memory structure for the streaming matrix (i.e., the tile filler STR and the tile reader STR) and a single controller to orchestrate the memory structure for C (i.e., the tile writer C).
Fig.~\ref{fig:memory_controllers_code} shows the code of these unified memory controllers. 

\section{Experimental Methodology}  
\label{section:methodology}

\textbf{Simulation Infrastructure}: For a detailed evaluation of \AcceleratorName, we have implemented a  cycle-level microarchitectural simulator of all on-chip components of our accelerator by leveraging the STONNE framework~\cite{stonne}\footnote{We plan to open-source the framework after the revision of the paper.}. To faithfully model the whole memory hierarchy including an HBM 2.0 off-chip DRAM, we have connected the simulated accelerator to the Structural Simulation Toolkit~\cite{SST2010}.
Table~\ref{table:accelerator_parameters} shows the main parameters of the architecture we have configured for the rest of the evaluation.  We compare our results against three state-of-the-art accelerators: SIGMA-like as an example of an \innerprod~accelerator, Sparch-like as an example of an \outerprod~accelerator and GAMMA-like as an example of a \gustavsons~accelerator.

We use the term \textit{-like} in GAMMA\textit{-like}, SIGMA\textit{-like} and SpArch\textit{-like} to reflect the fact that we capture their most relevant characteristics (i.e., their essence) in our simulator in a fair and normalized fashion. Specifically, we focus on the dataflow, which is a critical part for the efficiency of the accelerator; the DN, MN and RN components, which define the accelerator size and bandwidth; and the on-chip memory structures, which determine the capacity of the accelerator to store data close to the processing elements. 
We note that these are given extra on-chip area as appropriate. The main sources of efficiency in SIGMA, SpArch and GAMMA are the FAN reduction network, merge network and the fiber cache respectively rather than a specifically engineered design-point. Thus, we believe that the comparison against the key features of these designs captured by SIGMA-like, SpArch-like and GAMMA-like is justifiable, since our aim is to establish the advantages of flexibility and our ability to achieve it without major area overhead rather than obtain a specifically engineered design point.

For the three accelerators, we model the same parameters presented in Table~\ref{table:accelerator_parameters}, and we only change the memory controllers to deliver the data in the proper order according to its dataflow. 
We also compare \AcceleratorName~against the implementation from Intel MKL~\cite{MKL} running on a 4-core 8-thread Intel(R) Core(TM) i5-7400 CPU @ 3.00~GHz. Each core implements a 128 KiB L1 cache, a 1 MiB L2  cache and a shared 6 MiB L3 cache. 
We do not include GPU results because existing GPU SpMSpM implementations do not support sparse weights+activations natively~\cite{Zhu2020,SparseRT}, thus performing similarly to CPU MKL as reported in~\cite{SpArch2020,zhang2021gamma}.

To demonstrate the benefits of \AcceleratorName, our evaluation methodology considers the following three different angles:

\begin{table}[t!]
\begin{footnotesize}
\begin{center}
{
\begin{tabular}{|c|c|}\hline
 \textbf{Parameter} & \textbf{Value}\\\hline
 \textit{Number of Multipliers} & 64 \\\hline
 \textit{Number of Adders} & 63 \\\hline
 \textit{Distribution bandwidth} & 16 elems/cycle\\\hline
 \textit{Reduction/Merging bandwidth} & 16 elems/cycle\\\hline
 \textit{Total Word Size (Value+Coordinate)}              & 32 bits\\\hline
 \textit{L1 Access Latency}        & 1 cycle\\\hline
 \textit{L1 STA FIFO Size}  &  256 bytes \\\hline 
 \textit{L1 STR cache Size} & 1MiB \\\hline
 \textit{L1 STR Cache Line Size} & 128 bytes \\\hline
 \textit{L1 STR Cache Associativity} & 16 \\\hline
 \textit{L1 STR Cache Number of Banks} & 16 \\\hline
 \textit{PSRAM}   & 256 KiB \\\hline
 \textit{DRAM size}  & 16 GiB \\\hline
 \textit{DRAM access time / Bandwidth} & 100 ns / 256 GB/s \\\hline
 
\end{tabular}
}
\end{center}
\end{footnotesize}
\vspace{-0.35cm}
\caption{Configuration parameters of \AcceleratorName.}
\label{table:accelerator_parameters}
\vspace{-0.2cm}
\end{table}

\textbf{End-to-End Evaluation}: To truly prove the performance benefits of \AcceleratorName, we have carried out end-to-end execution of complete DNN models (see Table~\ref{table:dnn_models}) in our simulated accelerators. These models are present in the MLPerf benchmark suite~\cite{reddi2019mlperf} and we take other models for completeness. As it may be appreciated, we consider very diverse DNN models in terms of number of layers and sizes. The matrices involved in the execution of each DNN layer range from 0.003 MiB up to 63.41 MiB (see average compressed sizes in Table~\ref{table:dnn_models}), thereby our evaluation is comprehensive as there are many situations where matrices cannot completely fit on chip (\AcceleratorName~uses a total of 1 MiB SRAM memory for storing input matrices).

\textbf{Layer-wise evaluation}: Since explaining the results requires delving into a finer-grained detail, we have selected 9 representative layers extracted from the execution of the DNN models. Table~\ref{table:dnn_layers} shows these layers together with the characteristics of each layer.
\begin{table}[t!]
\begin{footnotesize}
\begin{center}
{
\begin{tabular}{|c|c|c|c|c|c|c| }\hline
 \textbf{Layer}  & \textbf{M, N, K} 
 & \textbf{spA} & \textbf{spB} & \textbf{csA} & \textbf{csB} & \textbf{csC} \\\hline
 \textit{SQ5} & 64, 2916, 16
 & 68 & 11 & 1.2 & 162 & 728 \\\hline
 \textit{SQ11} & 128, 729, 32 & 70 & 10 & 4.8 & 82 & 364 \\\hline
 \textit{R4} & 256, 3136, 64 & 88 & 9 & 7.6 & 709 & 3136\\\hline
 \textit{R6} & 64, 2916, 576 & 89 & 53 & 16 & 3086 & 728\\\hline
 \textit{S-R3} & 64, 5329, 576 & 89 & 46 & 16 & 6422 & 1332 \\\hline
 \textit{V0} & 128, 12100, 576 & 90 & 61 & 29 & 21357 & 12321 \\\hline
 \textit{MB215} & 128, 8, 512 & 50 & 0 & 128 & 16 & 4 \\\hline
 \textit{V7} & 512, 144, 4608 & 90 & 94 & 921 & 177 & 288 \\\hline
 \textit{A2} & 384, 121, 1728 & 70 & 54 & 777 & 373 & 181 \\\hline
\end{tabular}
}
\end{center}
\end{footnotesize}
\vspace{-0.35cm}
\caption{Representative DNN layers selected for the evaluation. sp\{A,B\}=sparsity of matrix \{A,B\} (in \%), cs\{A,B,C\}=compressed size of matrix \{A,B,C\} (in KiB).}
\label{table:dnn_layers}
\vspace{-0.2cm}
\end{table}

\textbf{RTL results}: 
We implemented the main building blocks (i.e., the DN, MN, RN and the on-chip memory) of the accelerators considered in this work (shown in Table~\ref{table:accelerator_models}).
For an apples-to-apples comparison of overheads, the four architectures use the same tree topology for the DN, the same linear array of multipliers for the MN and vary the RN. For the SIGMA-like architecture, we utilize the FAN network~\cite{SIGMA2020} as the RN for flexible-sized reductions. For the Sparch-like and GAMMA-like architectures, we use a merger~\cite{SpArch2020,outerspace} to merge the partial sums produced after the multiplications. Finally, for \AcceleratorName~we utilize the unified MRN explained in Section~\ref{section:design}. 

\begin{table}[t!]
\begin{footnotesize}
\begin{center}
{
\begin{tabular}{|c|c|c|c|c|}\hline
 & \textbf{SIGMA-} 
 & \textbf{Sparch-} & \textbf{GAMMA-} & \textbf{Flexagon}\\
 & \textbf{like} & \textbf{like} & \textbf{like} & \\
 \hline
  \textbf{DN} & Tree  &  Tree  & Tree & Tree\\\hline
   \textbf{MN} & Linear  & Linear           & Linear & Linear\\\hline
 \textbf{RN}    & FAN  & MergerS        & MergerG & MRN\\\hline
\end{tabular}
}
\end{center}
\end{footnotesize}
\vspace{-0.35cm}
\caption{Main building blocks to model the SIGMA-like, Sparch-like, GAMMA-like and \AcceleratorName~ accelerators. DN=Distribution Network, RN=Reduction Network and MN= Multiplier Network.  
}
\label{table:accelerator_models}
\vspace{-0.25cm}
\end{table}

For synthesis, we use MAERI BSV~\cite{MaeriCode} to generate the 64-MS distribution network and the multiplier network.  In addition,  we have implemented in RTL a 64-wide merger and our MRN.  We use Synopsys Design Compiler and Cadence Innovus Implementation System for synthesis and place-and-route, respectively, using TSMC 28nm GP standard LVT library at 800 MHz. To obtain the area and power numbers of the memory structures, we have used CACTI 7.0~\cite{cacti70} for the same technology node and frequency. 

\section{Results}
\vspace{-0.3cm}
\label{section:evaluation}
\subsection{End-to-end results}

\begin{figure}[t!]
        \begin{center}
                \includegraphics[width=1.0\columnwidth]{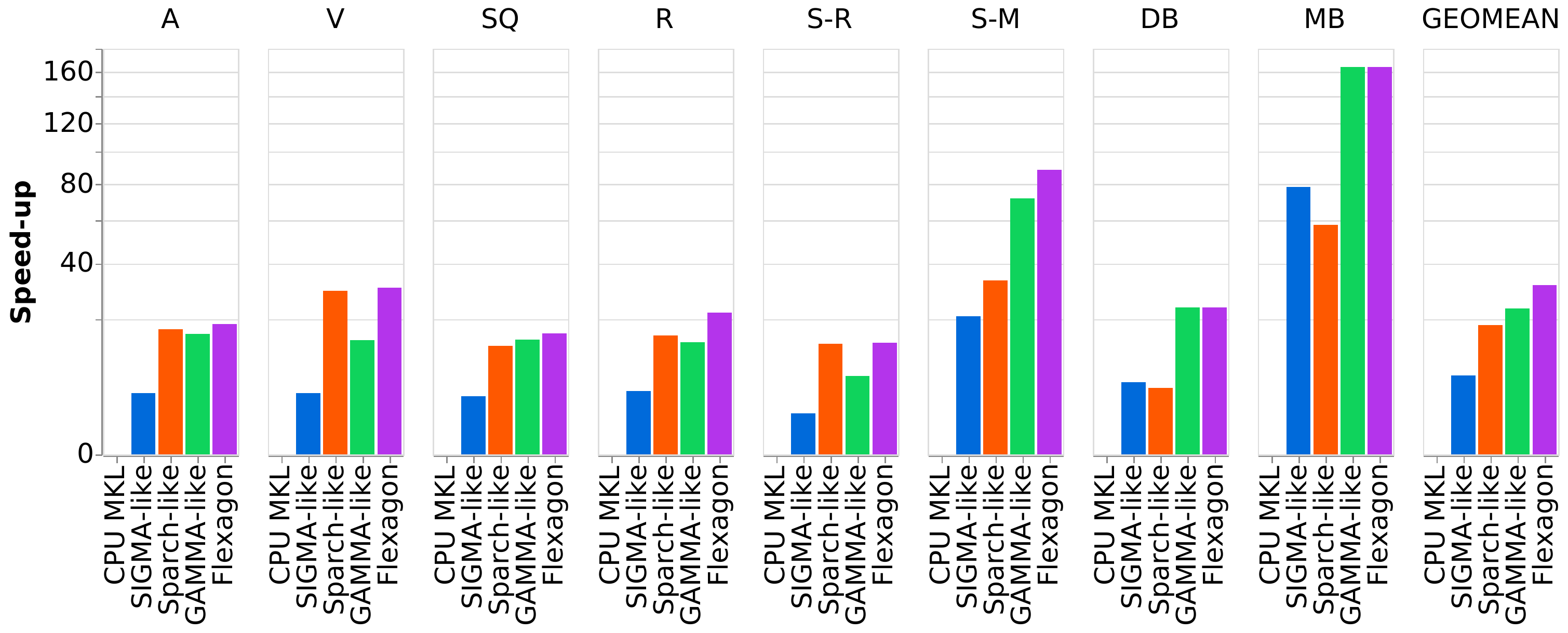}
        \end{center}
        \vspace{-0.40cm}
        \caption{Performance comparison between CPU MKL, SIGMA-like, Sparch-like, GAMMA-like and \AcceleratorName~architectures across the 8 DNN models (speed-up against SIGMA-like).}
        \label{fig:cycles_models}
        \vspace{-0.2cm}
\end{figure}

Figure~\ref{fig:cycles_models} compares the performance obtained with the CPU MKL, the three contemporary fixed-dataflow accelerators (SIGMA-like, Sparch-like and GAMMA-like) and with Flexagon when running the 8 DNN models (speed-ups with respect to the results obtained with the CPU MKL). The total numbers of cycles for CPU MKL are reported in the last column of Table~\ref{table:dnn_models}.

The first observation is that there is no fixed-dataflow accelerator that can obtain the highest performance for all the 8 DNN models. In particular, for \textit{Alexnet} (\texttt{A}), \textit{VGG-16} (\texttt{V}), \textit{Resnets-50} (\texttt{R}) and \textit{SSD-Resnets} (\texttt{S-R}) the Sparch-like accelerator is 5.26$\times$ and 1.49$\times$ on average faster than the SIGMA-like and GAMMA-like architectures, respectively. Conversely, for \textit{Squeezenet} (\texttt{SQ}), \textit{SSD-Mobilenets} (\texttt{SM}), \textit{DistilBert} (\texttt{DB}) and \textit{MobileBert} (\texttt{MB}), the GAMMA-like accelerator obtains the best performance (average improvements of 3.28$\times$ and 2.41$\times$ against the SIGMA-like and Sparch-like, respectively). 


The second and most noteworthy observation is that \AcceleratorName~can outperform the other three fixed-dataflow accelerators in all cases, attaining average speed-ups of 4.59$\times$ (vs. SIGMA-like), 1.71$\times$ (vs. Sparch-like) and 1.35$\times$ (vs. GAMMA-like). This is due to the combination of its flexible interconnects, explicitly decoupled memory structures and unified memory controllers that enable using the most efficient dataflow for each layer. 

Finally, we observe that \AcceleratorName~significantly outperforms the CPU MKL as the hardware is specifically designed to perform the SpMSpM operation. Overall, we find that \AcceleratorName~obtains a speed-up of 31$\times$ on average (benefits from 13$\times$ up to 163$\times$ are observed).

\subsection{Layer-wise results}
\vspace{-0.3cm}
Detailing the reasons behind the benefit observed for some DNN models for a particular dataflow requires a deeper delve into every DNN layer execution. To make the study feasible (we run over a hundred of layers), next, we present a comprehensive study for a selected set of nine representative DNN layers (Table~\ref{table:dnn_layers}). These layers are chosen according to the dataflow from which they benefit the most --The first three layers in the table benefit from \innerprod~(\textit{SQ5}, \textit{SQ11} and \textit{R4}), the second ones from \outerprod~(\textit{R6}, \textit{S-R3} and \textit{V0}), and the third ones from \gustavsons~(\textit{MB215}, \textit{V7} and \textit{A2}).

\begin{figure}[t!]
        \begin{center}
                \includegraphics[width=1.0\columnwidth]{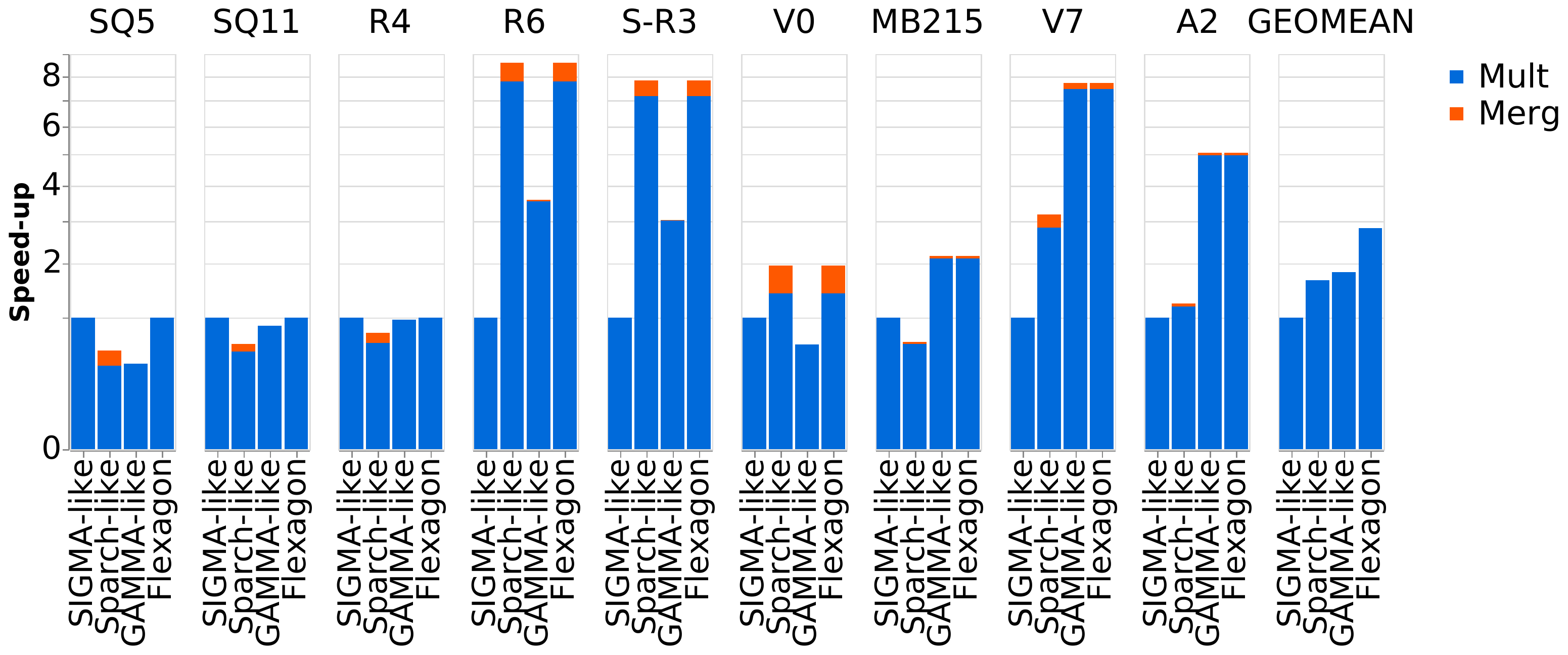}
        \end{center}
        \vspace{-0.40cm}
        \caption{Performance comparison between SIGMA-like, Sparch-like, GAMMA-like and \AcceleratorName~architectures across our 9 DNN layers (speed-up against the SIGMA-like one).}
        \label{fig:cycles_layers}
\end{figure}

Figure~\ref{fig:cycles_layers} shows a performance comparison running these selected layers using our simulated accelerators (again, speed-ups are computed with respect to SIGMA-like). 
%
%
As expected, as shown in the figure, in case of the first group of \innerprod-friendly layers, the SIGMA-like architecture obtains average speed-ups of 1.53$\times$ and 1.40$\times$ against the Sparch-like and the GAMMA-like architectures, respectively. 
The next three \outerprod-friendly layers (i.e., \textit{R6}, \textit{S-R3} and \textit{V0}), the Sparch-like architecture obtains an average increased performance of 5.07$\times$ and 2.66$\times$ against the SIGMA-like and GAMMA-like architectures. Finally, the last three \gustavsons-friendly layers, the best performance is obtained by the GAMMA-like architecture, experimenting 4.37$\times$ and 3.19$\times$ faster executions than the SIGMA-like and the Sparch-like architectures, respectively. More remarkable is that \AcceleratorName~beats all of them, always reaching the performance of the best case. 
Overall, by properly configuring the control logic of \AcceleratorName~according to the most suitable dataflow for each layer, our accelerator is able to attain 2.81$\times$, 1.69$\times$, and 1.55$\times$ speed-ups against the SIGMA-like, Sparch-like and GAMMA-like accelerators.

\begin{figure}[t!]
        \begin{center}
                \includegraphics[width=1.0\columnwidth]{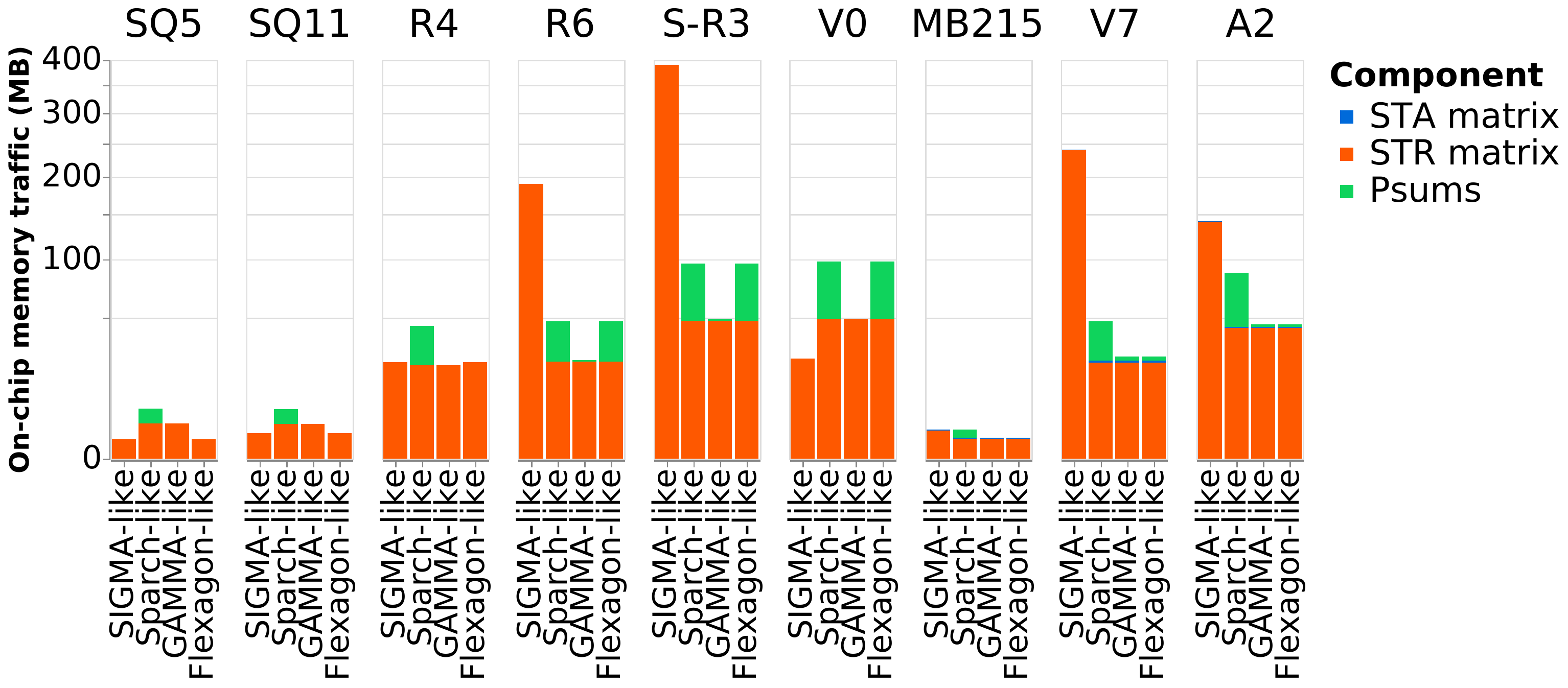}
        \end{center}
        \vspace{-0.40cm}
        \caption{Memory traffic (MB) that flows through the on-chip memory hierarchy for SIGMA-like, Sparch-like, GAMMA-like and \AcceleratorName~architectures across our 9 DNN layers.}
        \label{fig:memory_traffic_layers}
\end{figure}

\begin{figure}[t!]
        \begin{center}
                \includegraphics[width=1.0\columnwidth]{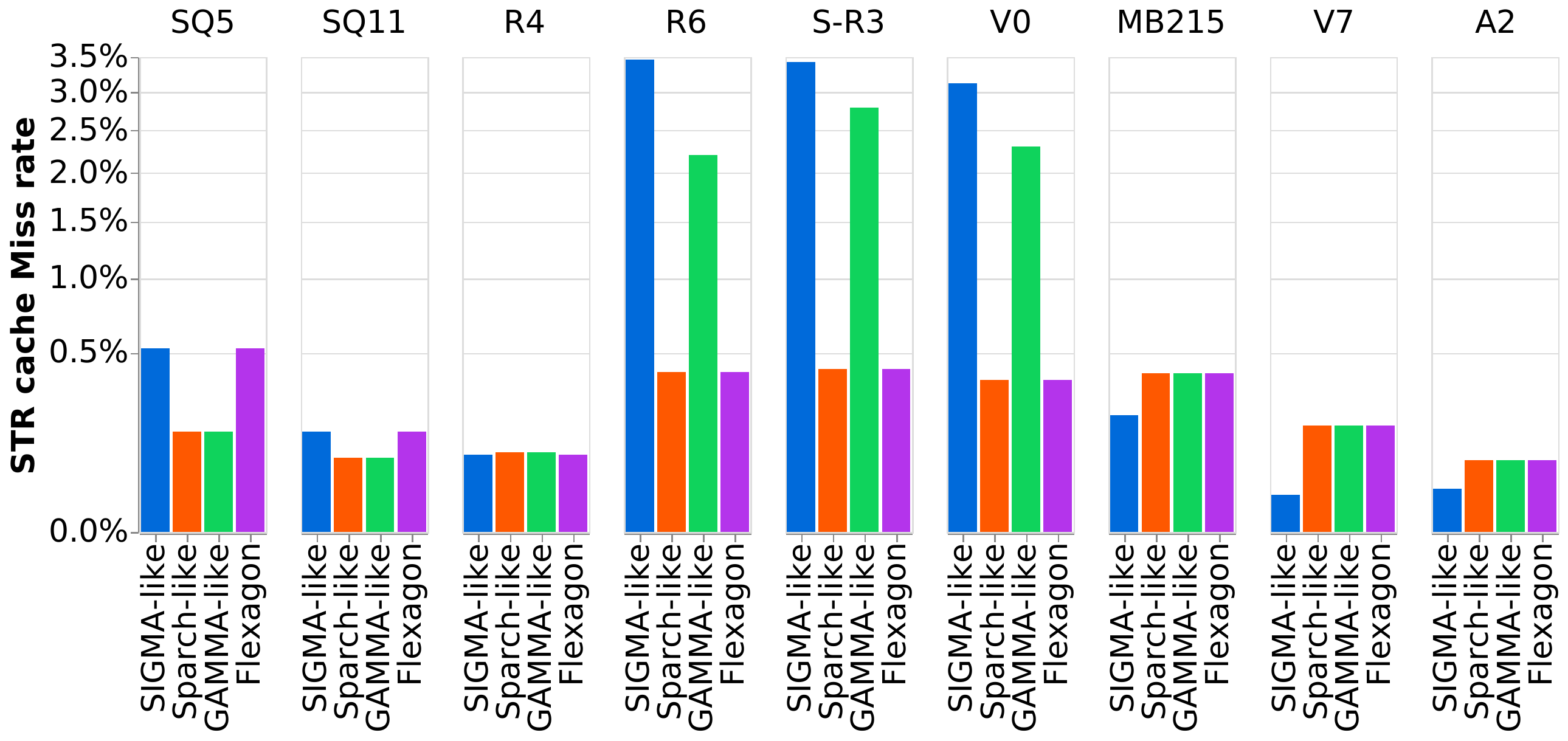}
        \end{center}
        \vspace{-0.4cm}
        \caption{STR cache miss rate for the SIGMA-like, Sparch-like, GAMMA-like and \AcceleratorName~architectures across 9 DNN layers.}
        \label{fig:missrate_layers}
        \vspace{-0.1cm}
\end{figure}

\begin{figure}[t!]
        \begin{center}
                \includegraphics[width=1.0\columnwidth]{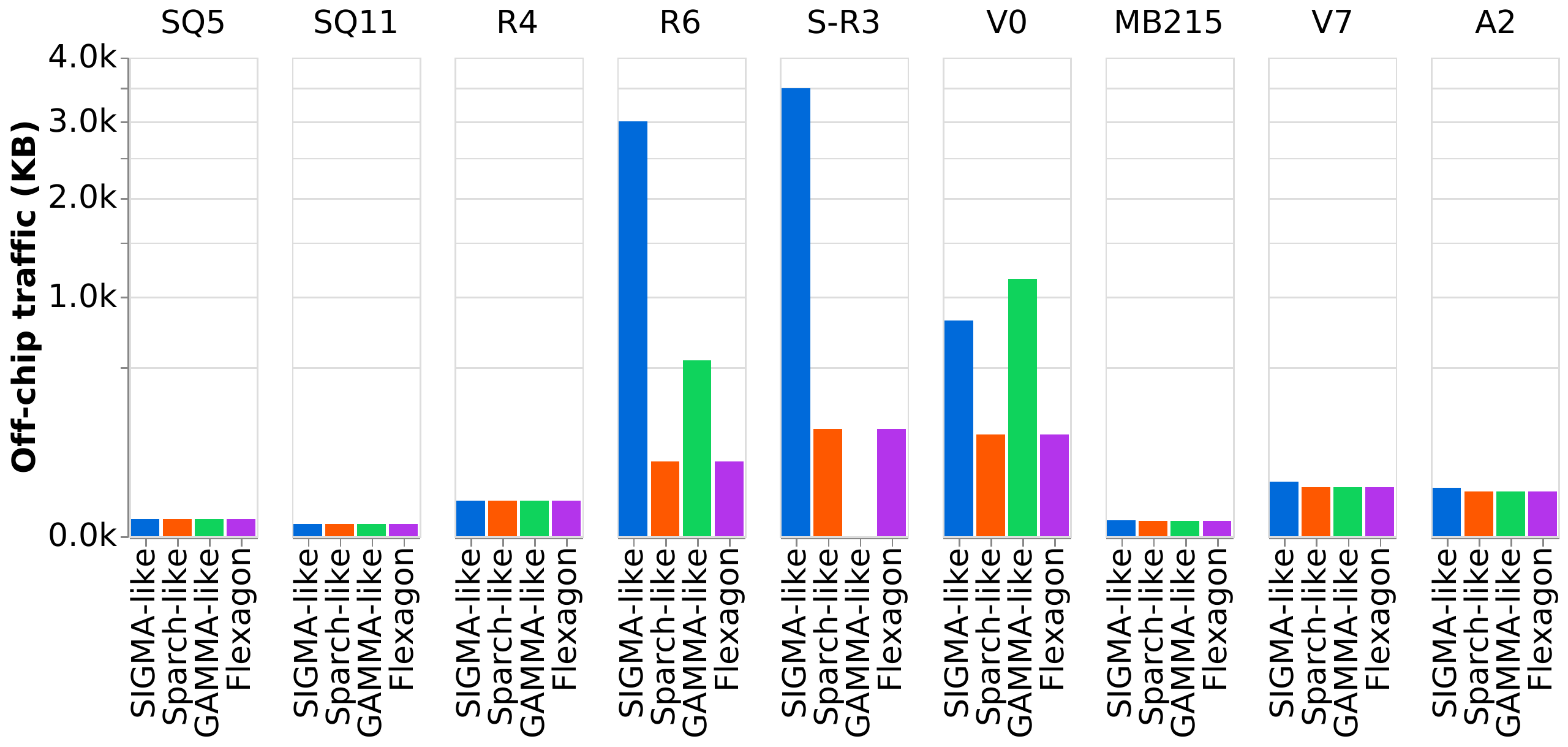}
        \end{center}
        \vspace{-0.40cm}
        \caption{Off-chip data traffic for the SIGMA-like, Sparch-like, GAMMA-like and \AcceleratorName~architectures across 9 DNN layers.}
        \label{fig:offchip_traffic}
        \vspace{-0.1cm}
\end{figure}

Figures~\ref{fig:memory_traffic_layers},~\ref{fig:missrate_layers} and~\ref{fig:offchip_traffic} help us understand these results. Specifically, Figure~\ref{fig:memory_traffic_layers} shows the amount of on-chip memory traffic (expressed in MBs) that relays between our on-chip memory hierarchy (i.e., the reads from the STA FIFO and from the STR cache and the reads/writes from/to the~\SramCName) and the distribution network after running the SIGMA-like, Sparch-like, GAMMA-like and \AcceleratorName~architectures across our nine DNN layers. Figure~\ref{fig:missrate_layers} plots the cache miss rate of the STR cache after running the layers, and Figure~\ref{fig:offchip_traffic} shows the amount of off-chip traffic (expressed in KBs) that in consequence,  flows between this STR cache and the DRAM.

The first observation that we would like to make from Figure~\ref{fig:memory_traffic_layers} is the negligible traffic that is fetched from the memory structure for the STA matrix (inappreciable fractions of the bars in blue color). This is basically due to the fact that the stationary data is kept stationary in the multipliers once it is read for the rest of the execution, as it is explained in Section~\ref{section:design}. For this reason, this memory structure does not have a significant impact on the final performance of the executions regardless of the dataflow that is configured.  
In contrast, the amount of traffic required to fill the structure for the STR matrix and the \SramCName heavily varies layer by layer and across dataflows (fractions of the bars in orange and green colors respectively), hence determining the final performance of the layer execution.

Since the \innerprod~ dataflow does not require to merge the partial sums as they are internally accumulated (observe the number of partial sums sent to the \SramCName for the SIGMA-like architecture is always 0) this dataflow obtains the best performance. An outlier for this behaviour is observed for the \textit{V0} layer. Here, the traffic generated for the STR matrix in the SIGMA-like architecture is lower than the traffic generated in the Sparch-like and GAMMA-like architectures. However, this workload experiences higher runtime. The reason of this is the large size of the matrix B (21.3 MiB) which causes that it has to be reloaded several times, experimenting a L1 miss rate of 3.13\% (see Figure~\ref{fig:missrate_layers}), significantly higher than the L1 miss rates obtained for the Sparch-like and GAMMA-like architectures (i.e., 0.36\% and 2.30\%) which translates into increased off-chip memory traffic (see Figure~\ref{fig:offchip_traffic}). This higher traffic provokes that the multiplying phase takes longer for the SIGMA-like architecture than for both the multiplying and merging phase for the Sparch-like architecture. When the number of intersections is low, the SIGMA-like architecture experiments higher number of cycles overheads due to this architecture accesses to many more data elements. This is also observed in the six layers that do not benefit from the SIGMA-like architecture (i.e., \textit{R6}, \textit{S-R3}, \textit{V0}, \textit{MB215}, \textit{V7} and \textit{A2}), experiencing on average 5.68$\times$ and 2.27$\times$ higher on-chip traffic than the Sparch-like and GAMMA-like architectures.

On the other hand, out of these six layers, the main difference of performance that defines them comes from the size of the matrix B. The second group of layers (i.e., \textit{R6}, \textit{S-R3} and \textit{V0}) that benefit from the Sparch-like architecture have a large size of matrix B (see Table~\ref{table:dnn_layers}). This implies that the GAMMA-like architecture cannot fit the rows of B entirely in the memory structure for the STR matrix, causing higher L1 miss rates. Observe the average L1 miss rate (see Figure~\ref{fig:missrate_layers}) experimented in the execution of these three layers is 0.39\% for the Sparch-like architecture and 2.43\% for the GAMMA-like architecture. This translates into 6.25$\times$ more traffic for GAMMA which causes the degradation in performance. 

In the last group of layers (i.e., \textit{MB215}, \textit{V7} and \textit{A2}) the size of matrices B are much smaller (up to 373KB as observed in Table~\ref{table:dnn_layers}) and therefore both Sparch-like and GAMMA-like architectures experience the same L1 miss rates and off-chip data traffic. In this scenario, the GAMMA-like architecture is more efficient as it is able to compute the merging phase and the merging phase at the same time --Observe the orange bar for the GAMMA-like cases in the Figure~\ref{fig:cycles_layers} is not significant as the merge phase is computed in parallel within the multiplying phase (i.e., blue bar).



\subsection{RTL results}
\vspace{-0.3cm}
\label{section:rtl_results}

\begin{table}[t!]
\begin{scriptsize}
\begin{center}
{
\begin{tabular}{|c|c|c|c|c|}\hline
 \textbf{Component} & \textbf{SIGMA-} & \textbf{Sparch-} & \textbf{GAMMA-}  & \textbf{Flexagon} \\
  & \textbf{like} & \textbf{like} & \textbf{like} & \\
 \hline
 \multicolumn{5}{|c|}{\textbf{Area Results}}\\\hline
 \textbf{DN (mm$^2$)} & 0.04 & 0.04 & 0.04  & 0.04 \\\hline
 \textbf{MN (mm$^2$)} & 0.07 & 0.07 & 0.07  & 0.07 \\\hline
 \textbf{RN (mm$^2$)} & 0.17 & 0.07 & 0.07  & 0.21 \\\hline
 \textbf{Cache (mm$^2$)} & 3.93 & 3.93 & 3.93  & 3.93 \\\hline
 \textbf{PSRAM (mm$^2$)} & - & 1.03 & 0.51  & 1.03 \\\hline
 \textbf{Total (mm$^2$)} & 4.21 & 5.14 & 4.62  & 5.28 \\\hline
 \multicolumn{5}{|c|}{\textbf{Power Results}}\\\hline
  \textbf{DN (mW)} & 2.18 & 2.18 & 2.18  & 2.18 \\\hline
 \textbf{MN (mW)} & 3.29 & 3.29 & 3.29  & 3.29 \\\hline
 \textbf{RN (mW)} & 248 & 64.48 & 64.48  & 312 \\\hline
 \textbf{Cache (mW)} & 2142 & 2142 & 2142  & 2142 \\\hline
 \textbf{PSRAM (mW)} & - & 538 & 269 &  538 \\\hline
 \textbf{Total (mW)} & 2396 & 2750 & 2481  & 2998 \\\hline
\end{tabular}
}
\end{center}
\end{scriptsize}
\vspace{-0.3cm}
\caption{Post-layout area and power obtained for SIGMA-like Sparch-like, GAMMA-like and Flexagon accelerators.}
\label{table:area}
\vspace{-0.15cm}
\end{table}

Table~\ref{table:area} shows a  breakdown  of  the  total  amount  of  area (mm$^2$) and power (mW) obtained for the 64-MS SIGMA-like, Sparch-like, GAMMA-like and \AcceleratorName~accelerators. For each case, we show the results for the main architectural components:  Distribution Network (DN), Multiplier  Network (MN), Reduction/Merger Network (RN), the cache structure for the streaming matrix (Cache) and the \SramCName. 

In terms of area, we observe that \AcceleratorName~introduces an overhead of 25\%, 3\% and 14\% with respect to the area of the SIGMA-like, Sparch-like and GAMMA-like accelerators, respectively. As we can see, the area of the four accelerators is mostly dominated by the memory structures. Specifically, we observe that the cache for the streaming matrix represents a 93\%, 76\%, 85\% and 74\% of the total amount of area for the SIGMA-like, Sparch-like, GAMMA-like and \AcceleratorName~architectures, respectively. Besides, the area of the \SramCName represents a  20\%, 11\% and 19\% with respect to the Sparch-like, GAMMA-like and \AcceleratorName~accelerators, respectively. Since the SIGMA-like architecture employs an \innerprod~dataflow, this accelerator does not need this structure, which explains the reason of having the lowest area. Also, the area of the \SramCName is the GAMMA-like accelerator is half the area in the Sparch-like and \AcceleratorName~accelerators as it requires to store less partial sums, which explains the area reduction. Obviously, \AcceleratorName~needs support for the worst-case \outerprod~dataflow and needs the highest \SramCName overhead.  Finally, note that our MRN is 28\% and 128\% larger than the area of the FAN and the merger, but this does not translates into high overall overhead as the MRN takes only a 4\% out of the total area for \AcceleratorName.

 \begin{figure}[t!]
        \begin{center}
                \includegraphics[width=0.65\columnwidth]{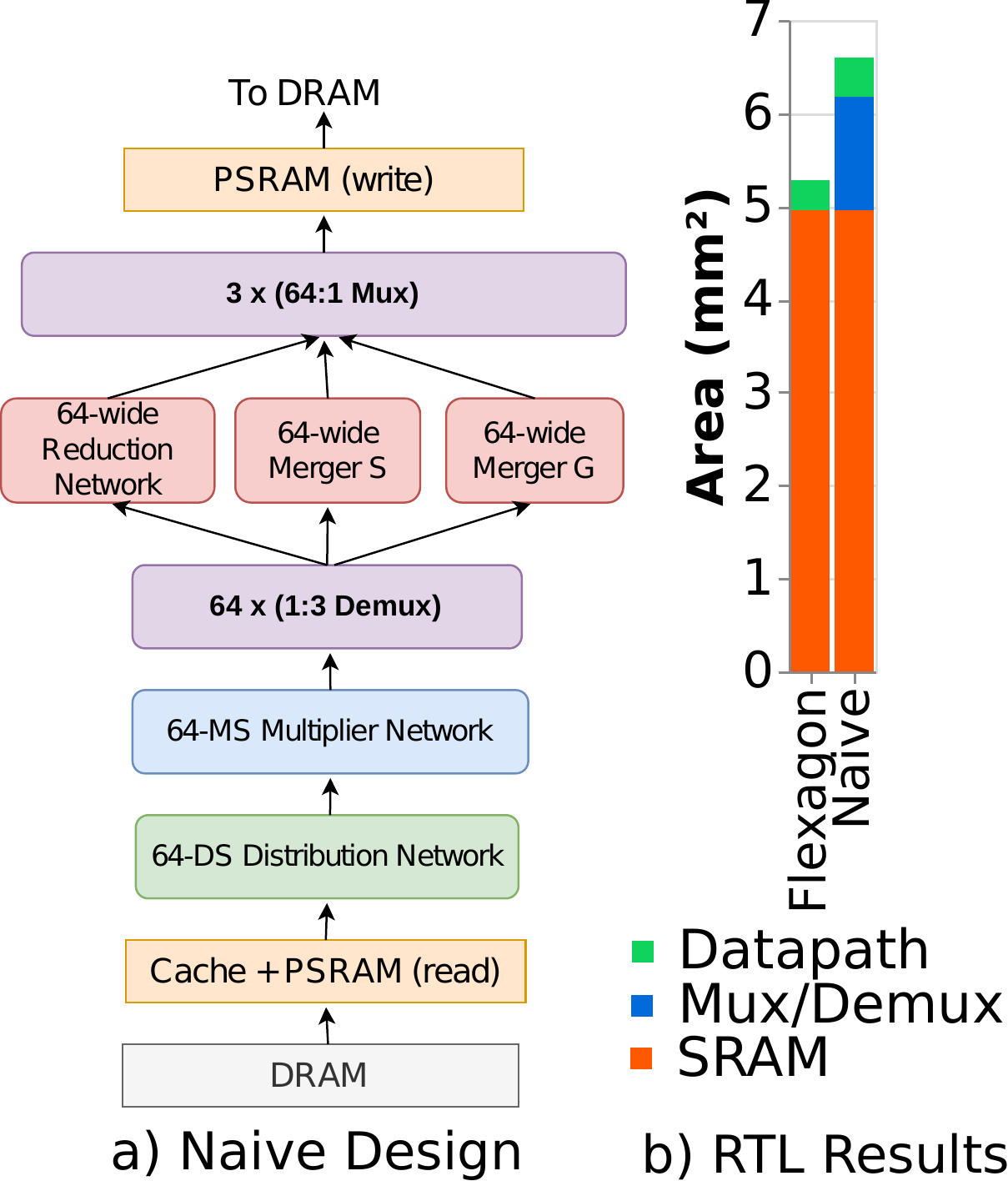}
        \end{center}
        \vspace{-0.40cm}
        \caption{a) High-level overview of a non-unified naive design. b) Area comparison between Flexagon and the naive design.}
        \label{fig:benefits_unification}
\end{figure}

Figure~\ref{fig:benefits_unification} proves the area benefits of unifying the RN and the merger into a single network (MRN). To do so, we have sketched a 64-MS naive accelerator design similar to Flexagon, but utilizing separate networks for each dataflow (see Figure~\ref{fig:benefits_unification}a).  We use the term \textit{naive} here to emphasize the fact that the design simply replicates the reduction network 3 times (one per each dataflow).  As it may be seen, the reduction and merger networks share the same multiplier and distribution networks as well as the same SRAM capacity. The design requires extra links, muxes and demuxes to connect the pieces. At the bottom side, the MN connects to three different networks, and therefore, requires 64 (1:3) demultiplexers. At the top side, each node from the merger and reduction network has to be connected to memory requiring 3 costly (64:1) multiplexers and connections. Figure~\ref{fig:benefits_unification}b shows the inefficiencies of this naive design. As we can see, the three separate networks (i.e., RNs and mergers) introduce an area overhead of just 2\% as the designs are dominated by the SRAM area (e.g., 74\% of area for Flexagon). The significant area penalty introduced by the naive design comes from the extra multiplexers, demultiplexers and corresponding connections, introducing an area overhead of 25\% over Flexagon. Note that in larger configurations (i.e., greater number of multipliers) this area overhead would even increase.

\begin{figure}[t!]
        \begin{center}
                \includegraphics[width=1.0\columnwidth]{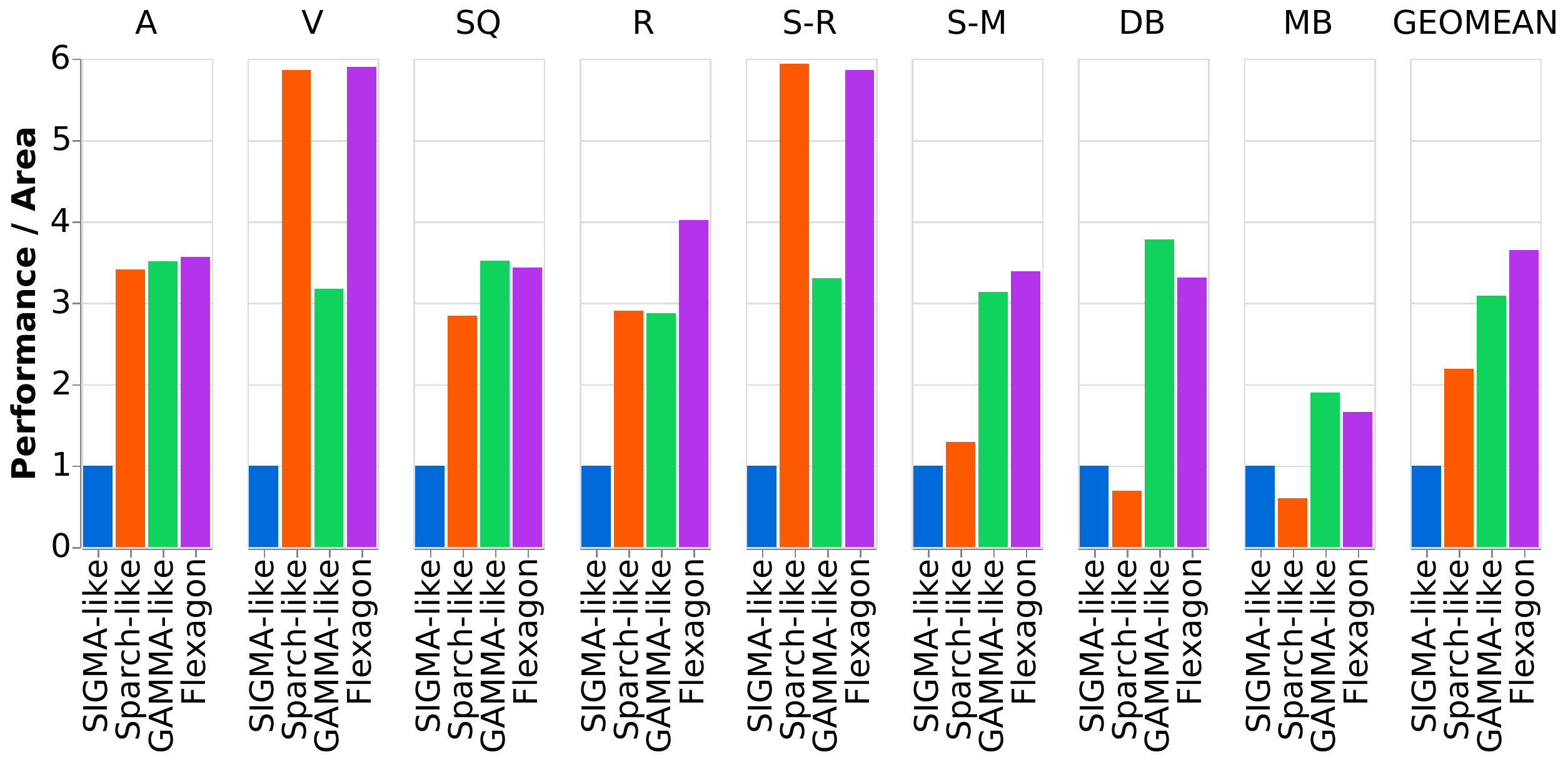}
        \end{center}
        \vspace{-0.40cm}
        \caption{Performance/Area obtained after running the SIGMA-like, Sparch-like, GAMMA-like and \AcceleratorName~architectures across our 8 DNN models.}
        \label{fig:perfarea_models}
        \vspace{-0.1cm}
\end{figure}

In terms of power, we observe the same trends. We find that the \AcceleratorName~accelerator consumes 25\%, 9\% and 21\% more power than the SIGMA-like, Sparch-like and GAMMA-like accelerators. The slightly higher overhead of \AcceleratorName~against the aforementioned area results comes mostly from the  Merger/RN as this module represents a larger fraction of total consumption (10\%, 2.34\%, 2.60\% and 10.41\% out of the SIGMA-like, Sparch-like, GAMMA-like and flexagon accelerators are observed, respectively). This, together with the fact that the MRN consumes 25\% and 284\% more than the FAN RN and the merger, explains the results. In spite of the overhead introduced, in Figure~\ref{fig:perfarea_models} we illustrate that \AcceleratorName~is still more performance/area efficient. Specifically, we consider both achieved speed-ups and area requirements of each design. The area requirements are normalized with respect to the SIGMA-like case, which is also the reference for the calculation of the speed-ups. Note that the NLP models like \textit{MobileBert} (\textit{MB}) and \textit{DistilBert} (\textit{DB}) achieves a better efficiency with the GAMMA-like accelerator. Nevertheless, this is due to as explained before, most of the layers (84\% in DistilBert (\textit{DB}) and 100\% in MobileBert (\textit{MB})) for these models work better with the \texttt{Gustavson} dataflow, making the area overhead introduced by the \AcceleratorName~accelerator unnecessary.  Consequently, we can clearly see that, overall, \AcceleratorName~reaches the best compromise between performance and area consumption (the higher Speed-up/Area values). In comparison, we find that, on average, our accelerator obtains 18\%, 67\% and 265\% better performance/area efficiency across the execution of our 8 DNN models with respect to the GAMMA-like, Sparch-like and SIGMA-like accelerators. This makes \AcceleratorName~the best candidate for running heterogeneous sparse DNN workloads. 

\section{Related work}  

\vspace{-0.3cm}
\textbf{Sparse DNN Accelerators:}
Sparse matrix multiplications have been prime targets of acceleration for AI and HPC workloads. Several sparse DNN accelerators have been proposed for SpMM, SpGEMM and Sparse convolution ~\cite{han-isca2016,outerspace,hegde2019extensor,srivastava2020matraptor,eyerissv2, lee2018stitch, gondimalla2019sparten, SIGMA2020, kanellopoulos2019smash,SCNN2017}. These accelerators have support for sparse execution via compression of one or both operands into formats like CSR, CSC, bitmap, CSF etc. This reduces the memory footprint and the number of multiplications.
%
As Table~\ref{table:related} shows, prior sparse accelerators have picked either one of \innerprod, \outerprod~and \gustavsons (row-wise product) dataflows. We show that flexibility to support multiple dataflows is beneficial for performance and performance per area.

\textbf{Frameworks for flexible accelerators:}
Prior works in the direction of flexibility include hardware widgets and design-space exploration tools for CGRAs.
MINT~\cite{mint} is a format converter widget that supports multiple sparse formats. Prior works Garg et al.~\cite{garg2021understanding}, coSPARSE~\cite{cosparse} and SparseAdapt~\cite{sparseadapt} propose frameworks for efficient sparse execution on CGRAs. However, to the best of our knowledge, this is the first work that proposes an accelerator for Sparse DNNs which exploits all the three dataflows.



\section{Conclusion}
\label{section:conclusions}
\vspace{-0.3cm}
This work proposes \AcceleratorName, the first SpMSpM accelerator design that offers~\innerprod, \outerprod~and~\gustavsons~dataflows on a homogeneous hardware substrate. \AcceleratorName~ revolves around a novel tree-based network (MRN) that supports both reduction of dot products and merging of partial sums, and a special L1 on-chip memory organization, specifically tailored to the different access characteristics of the input and output compressed matrices. By using the dataflow that best matches the characteristics of each DNN layer, we show that \AcceleratorName~brings significant improvements in performance/area efficiency over SOTA fixed-dataflow sparse
accelerators.

\bibliographystyle{plain} 
\bibliography{flexagon-distrib}



\end{document}